\documentclass[twocolumn]{aastex631}

\usepackage{color}
\usepackage{graphicx}
\usepackage{subfigure}
\usepackage[utf8]{inputenc}
\usepackage{amsmath}
\usepackage{amssymb}
\usepackage{physics}
\usepackage{hyperref}

\newcommand{\diff}{\mathrm{d}}
\newcommand{\vect}[1]{\boldsymbol{#1}}

\submitjournal{Accepted for publication by ApJ Letters}

\begin{document}

\title{Cross-Field Diffusion Effects on Particle Transport in a Solar Coronal Flux Rope}

\correspondingauthor{Edin Husidic}
\email{edin.husidic@kuleuven.be}

\author[0000-0002-1349-3663]{Edin Husidic}
\affiliation{Centre for mathematical Plasma Astrophysics, Department of Mathematics, \\ 
KU Leuven, Celestijnenlaan 200B, 3001 Leuven, Belgium}
\affiliation{Department of Physics and Astronomy, University of Turku, 20014 Turku, Finland}

\author[0000-0001-6344-6956]{Nicolas Wijsen}
\affiliation{Centre for mathematical Plasma Astrophysics, Department of Mathematics, \\ 
KU Leuven, Celestijnenlaan 200B, 3001 Leuven, Belgium}

\author[0000-0002-4014-1815]{Luis Linan}
\affiliation{Centre for mathematical Plasma Astrophysics, Department of Mathematics, \\ 
KU Leuven, Celestijnenlaan 200B, 3001 Leuven, Belgium}

\author[0000-0003-0874-2669]{Michaela Brchnelova}
\affiliation{Centre for mathematical Plasma Astrophysics, Department of Mathematics, \\ 
KU Leuven, Celestijnenlaan 200B, 3001 Leuven, Belgium}

\author[0000-0002-3298-2067]{Rami Vainio}
\affiliation{Department of Physics and Astronomy, University of Turku, 20014 Turku, Finland}

\author[0000-0002-1743-0651]{Stefaan Poedts}
\affiliation{Centre for mathematical Plasma Astrophysics, Department of Mathematics, \\ 
KU Leuven, Celestijnenlaan 200B, 3001 Leuven, Belgium}
\affiliation{Institute of Physics, University of Maria Curie-Sk{\l}odowska, ul.\ Radziszewskiego 10, 820-031 Lublin, Poland}

\received{August 22nd, 2024}
\revised{September 15th, 2024}
\accepted{October 31st, 2024}

\begin{abstract}

Solar energetic particles (SEPs) associated with solar flares and coronal mass ejections (CMEs) are key agents of space weather phenomena, posing severe threats to spacecraft and astronauts. Recent observations by Parker Solar Probe (PSP) indicate that the magnetic flux ropes of a CME can trap energetic particles and act as barriers, preventing other particles from crossing. In this paper, we introduce the novel COCONUT+PARADISE model to investigate the confinement of energetic particles within a flux rope and the effects of cross-field diffusion (CFD) on particle transport in the solar corona, particularly in the presence of a CME. Using the global magnetohydrodynamic coronal model COCONUT, we generate background configurations containing a CME modeled as a Titov--D\'{e}moulin flux rope (TDFR). We then utilize the particle transport code PARADISE to inject monoenergetic 100\,keV protons inside one of the TDFR legs near its footpoint and evolve the particles through the COCONUT backgrounds. To study CFD, we employ two different approaches regarding the perpendicular proton mean free path (MFP): a constant MFP and a Larmor radius-dependent MFP. We contrast these results with those obtained without CFD. While particles remain fully trapped within the TDFR without CFD, we find that even relatively small perpendicular MFP values allow particles on the outer layers to escape. In contrast, the initially interior trapped particles stay largely confined. Finally, we highlight how our model and this paper's results are relevant for future research on particle acceleration and transport in an extended domain encompassing both the corona and inner heliosphere.

\end{abstract}


\section{Introduction} \label{sec:introduction}

The Sun emits a continuous stream of charged particles filling the heliosphere. Streaming through this ambient solar wind are so-called solar energetic particles (SEPs) consisting of electrons, protons and ions, which are accelerated during solar eruptive events such as solar flares and at the fronts of CME-driven shock waves \citep{Desai-Giacalone-2016, Reames-2017}. When directed towards Earth, SEPs pose a threat to satellites and astronauts in the near-Earth environment \citep{Vainio-etal-2009,Gopalswamy-2018}, as well as to ground-based technology on Earth \citep{Schrijver-etal-2014, Schrijver-etal-2015}. As a result, an ongoing effort exists to understand and predict these space weather events.

In-situ measurements by the Parker Solar Probe \citep[PSP;][]{Fox-etal-2016} within the solar corona have opened new opportunities to examine the interaction between magnetic flux ropes and SEPs. In this context, the massive SEP event of 2022 September 5, when PSP was at a radial distance of about $15\,R_\odot$, has sparked great interest \citep[e.g.,][]{Paouris-etal-2023, Trotta-etal-2024}. By examining energetic proton populations, \cite{Cohen-etal-2024} found that PSP recorded a significant intensity drop at the shock, followed by a rapid increase in intensity by multiple orders of magnitude as PSP crossed the CME flank and entered a region of closed magnetic field lines, identified as a magnetic cloud. Additionally, the authors highlight the confinement of the protons within the cloud and occasional anisotropies observed in the proton distribution inside the cloud. In another study, using data from PSP orbit 5 in 2020 at radial distances from 0.45\,au to 0.2\,au, \cite{Pecora-etal-2021} showed that flux ropes act as boundaries, separating trapped particle populations within the flux rope from those moving outside of it. \cite{Schwadron-etal-2024} reached similar conclusions when they investigated PSP observations from March 2022 at a radial distance of 0.2\,au, when PSP passed through the flank of a CME. They found energetic particle populations trapped within flux tubes, with the edges of the flux tubes acting as barriers to other particles. So-called Forbush decreases (FD; \citealt{Cane-2000, Belov-2008}) support the view that magnetic flux ropes can act as barriers for energetic particles. Observations of galactic cosmic ray (GCR) fluxes often show sudden two-step decreases that are attributed, among other causes, to passing CMEs. The first step in the decay of GCR fluxes is associated with the CME-driven shock and the turbulent sheath between the shock and the driver. The second, faster step is then caused by isolated magnetic flux ropes blocking the particles. Recently, \cite{Benella-etal-2020} and \cite{Laitinen-Dala-2021} performed full-orbit particle simulations and showed that isolated flux ropes significantly hinder GCRs from penetrating the CME, while GCRs could enter the CME via the x-point. Besides the recent PSP observations, type IV radio bursts are another indicator for energetic particles trapped within flux ropes \citep{Morosan-etal-2019}. Among the proposed explanations for these phenomena is cyclotron emission by electrons trapped inside CME loops \citep{Bastian-2007}. 

Numerical models can prove invaluable for testing particle transport by simulating SEP events and helping to understand the underlying mechanisms. Over the last two decades, efforts have been made to develop and improve different simulation tools, usually combinations of a magnetohydrodynamic (MHD) model that provides coronal solar wind background configurations and an energetic particle transport code. One example is the Energetic Particle Radiation Environment Module (EPREM; \citealt{Kozarev-etal-2010}) that utilizes input from CORona-HELiosphere (CORHEL; \citealt{Young-etal-2021}), both part of the Solar particle event Threat Assessment Tool (STAT) software suite. Another recent model is the Multiple-Field-Line-Advection Model for Particle Acceleration (M-FLAMPA; \citealt{Borovikov-etal-2018}) that takes input from the Alfv\'{e}n wave-driven solar atmosphere model (AWSoM; \citealt{Sokolov-etal-2021}) to simulate particle acceleration and transport in the corona. Furthermore, iPATH (improved Particle Acceleration and Transport in the Heliosphere; \citealt{Hu-etal-2017,Ding-etaL-2024}) uses MHD simulations of CME-driven shocks to model SEP events upstream of these shock waves. 

To enhance our understanding of particle transport and acceleration in the corona, we introduce COCONUT+PARADISE as the most recent advancement of our particle transport model. COCONUT (COolfluid COroNal UnsTructured; \citealt{Perri-etal-2022, Perri-etal-2023, Kuzma-etal-2023}) is a three-dimensional (3D) coronal MHD model developed within the framework of the Computational Object-Oriented Libraries for Fluid Dynamics (COOLFluiD; \citealt{Lani-etal-2005, Lani-etal-2006}), to eventually replace the semi-empirical Wang-Sheeley-Arge model in the current EUFHORIA (European Heliospheric Forecasting Information Asset; \citealt{Pomoell-Poedts-2018}) implementation. It is coupled to PARADISE (PArticle Radiation Asset Directed at Interplanetary Space Exploration; \citealt{Wijsen-2020}) that propagates energetic particles through the COCONUT backgrounds. Previously, PARADISE has been used only at radial distances  $r > 0.1$\,au with the heliospheric MHD solar wind and CME evolution and propagation models EUHFORIA and Icarus \citep{Verbeke-etal-2022} in both observational \citep{Wijsen-etal-2021, Wijsen-etal-2022, Wijsen-etal-2023} and theoretical studies \citep{Wijsen-etal-2019a,Husidic-etal-2024, Niemela-etaL-2024}. Furthermore, the ongoing coupling of the coronal COCONUT model to the heliospheric EUHFORIA model will enable us to investigate particle acceleration and transport with PARADISE from the low corona up to 1\,au and beyond consistently.

To demonstrate the capability of our model and its potential for future work involving in-situ observations while ensuring consistency with PSP's data, we simulate the evolution of SEP distributions within a CME and illustrate how cross-field diffusion (CFD) may affect the confinement of particles in the embedded CME flux rope. Using COCONUT, we generate coronal background configurations from 1 to 21.5 solar radii (0.1\,au), containing a CME modeled as an unstable modified Titov-D\'{e}moulin flux rope (TDFR) \citep{Titov-Demoulin-1999, Titov-etal-2014}. Unlike static coronal magnetic loops, which typically feature a quasi-stationary, untwisted magnetic structure, the erupting TDFR creates a more complex magnetic topology with varying magnetic field strengths and curvatures throughout the structure, which affect the particle transport dynamics. PARADISE subsequently injects energetic particles inside the flux rope close to one of its footpoints and evolves the particles through the COCONUT snapshots. To address the effects of CFD on particle transport inside the corona, we apply two approaches for CFD coefficients with varying parameter settings: a constant perpendicular mean free path (MFP) and a perpendicular MFP proportional to the particle's Larmor radius. We compare these results against those obtained without CFD. 

The remaining part of this paper is structured as follows. A short description of the utilized models, COCONUT and PARADISE, and the utilized diffusion conditions are discussed in Sec.~\ref{sec:model}. In Sec.~\ref{sec:results}, the simulation results are presented, starting with simulations without CFD (Sec.~\ref{subsec:no_CFD}), followed by simulations utilizing CFD with a constant perpendicular MFP (Sec.~\ref{subsec:const_mfp_perp}) and a Larmor radius-dependent perpendicular MFP (Sec.~\ref{subsec:larmor}). Section~\ref{sec:summary} concludes the paper by summarizing our results and highlighting the potential of our model for future research by describing some of the ongoing efforts. In the appendix, technical details of COCONUT (Sec.~\ref{app:coconut}) and PARADISE (Sec.~\ref{app:paradise}) are presented. Furthermore, in Sec.~\ref{app:interpolation}, the unstructured COCONUT grid is compared to its interpolation to a structured grid, while possible issues regarding numerical diffusion are addressed in Sec.~\ref{app:num_diffusion}.

\section{Numerical Models} \label{sec:model}

We utilize the two models, COCONUT and PARADISE, to simulate particle transport in the solar corona. We use a setup in COCONUT similar to that described in \cite{Linan-etal-2023} to obtain the necessary background corona configurations. The solar wind was reconstructed from a magnetogram by the Helioseismic and Magnetic Imager (HMI) onboard the Solar Dynamics Observatory (SDO) on 2nd July 2019 (during solar minimum), which provided the inner boundary conditions for the MHD equations. A CME is inserted into the simulation at $t = 0$\,h, modeled using the modified analytical circular Titov-D\'{e}moulin flux rope (TDFR) model \citep{Titov-Demoulin-1999, Titov-etal-2014}. 
The CME corresponds to the case $\zeta = 12$ in \cite{Linan-etal-2023}, where $\zeta$ is a parameter in the equation for the ring current and any $\zeta > 1$ results in an unstable TDFR from the beginning of the simulation. The TDFR has an initial magnetic field strength of $B_\mathrm{TDFR,0} = 10.5$\,G (or $10.5\times 10^{5}$\,nT), and the eruption of the TDFR leads to an initial speed of $v_\mathrm{TDFR,0} = 827$\,km/s. It is initially placed at $\theta = 90^\circ$ and $\phi = 180^\circ$, while the centers of the two footpoints are offset by a distance $d = 0.15\,R_\odot$ from the solar surface. Their major radii are $0.3\,R_\odot$ and their minor radii $0.1\,R_\odot$, resulting in a polarity area of $4,839$\,Mm$^2$. The entire COCONUT domain consists of about 2 million prism-shaped cells that are arranged in concentric shells and increase in size with radial distance (see Sec.~\ref{app:coconut} and \citealt{Brchnelova-etal-2022} for details.) The output cadence by COCONUT is set to 289\,s ($\approx 5$\,min). 
For Sec.~\ref{sec:results}, we use the first $\sim 7$\,h of the COCONUT simulation until the TDFR reaches the outer boundary. 
The outer boundary for the transport simulations is assumed to be an open boundary and is placed at $21.5\,R_\odot$. However, to remove any possible outer boundary effects in COCONUT, we simulate the corona to $25\,R_\odot$, as recommended by \cite{Brchnelova-etal-2022}.

\begin{figure*}[t]
    \centering
    \subfigure{\includegraphics[width=0.31\textwidth]{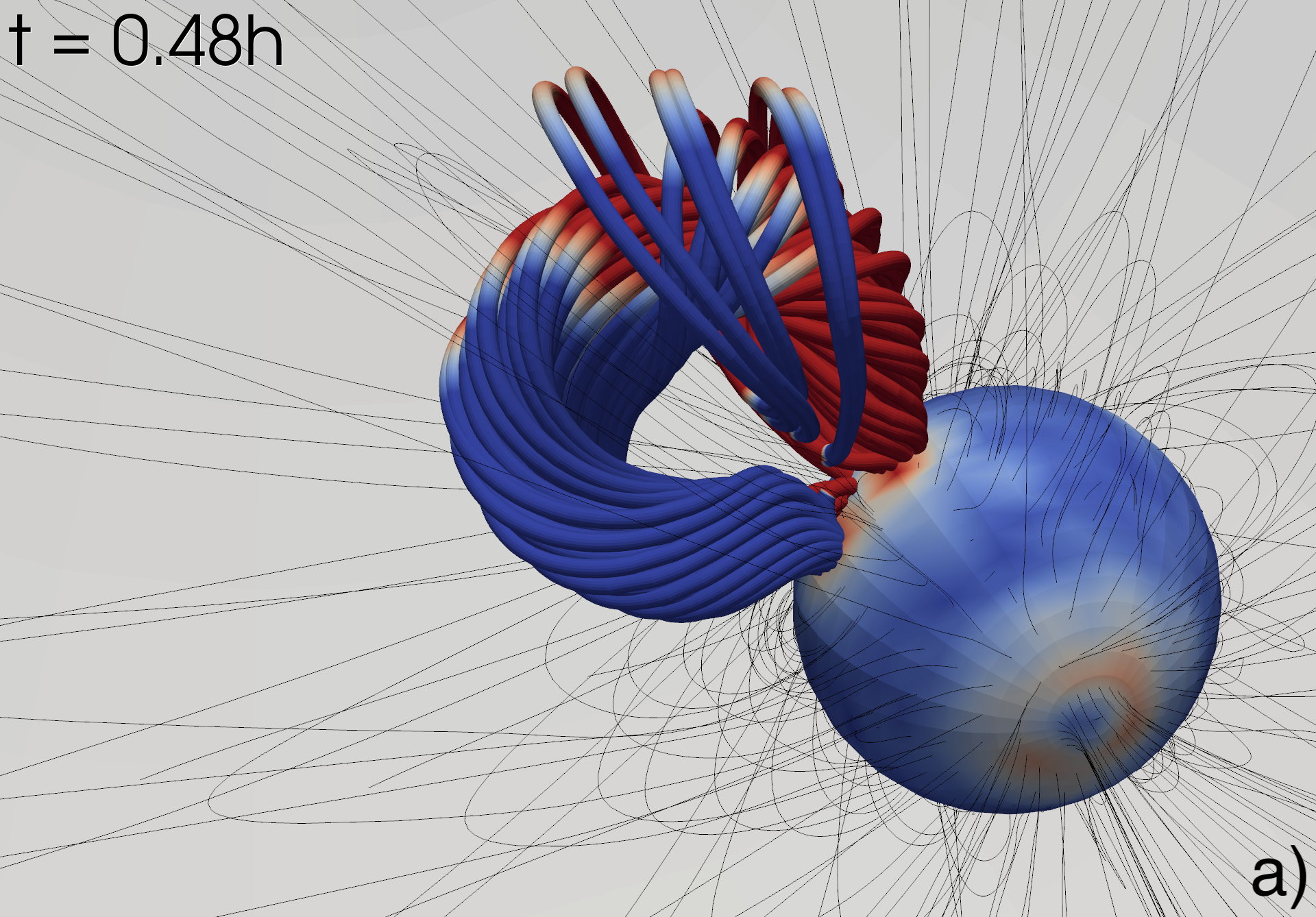}} \quad
    \subfigure{\includegraphics[width=0.31\textwidth]{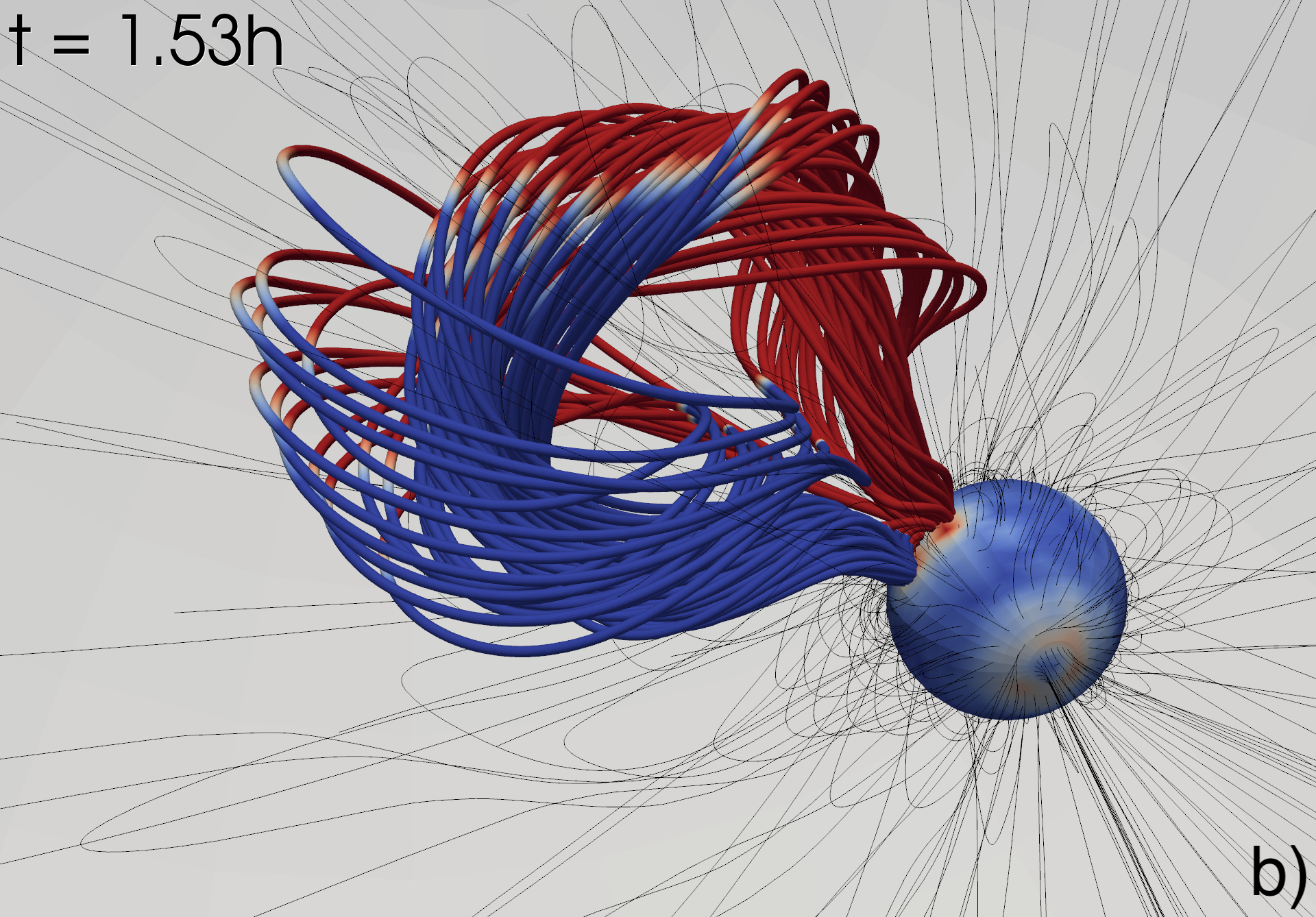}} \quad
    \subfigure{\includegraphics[width=0.31\textwidth]{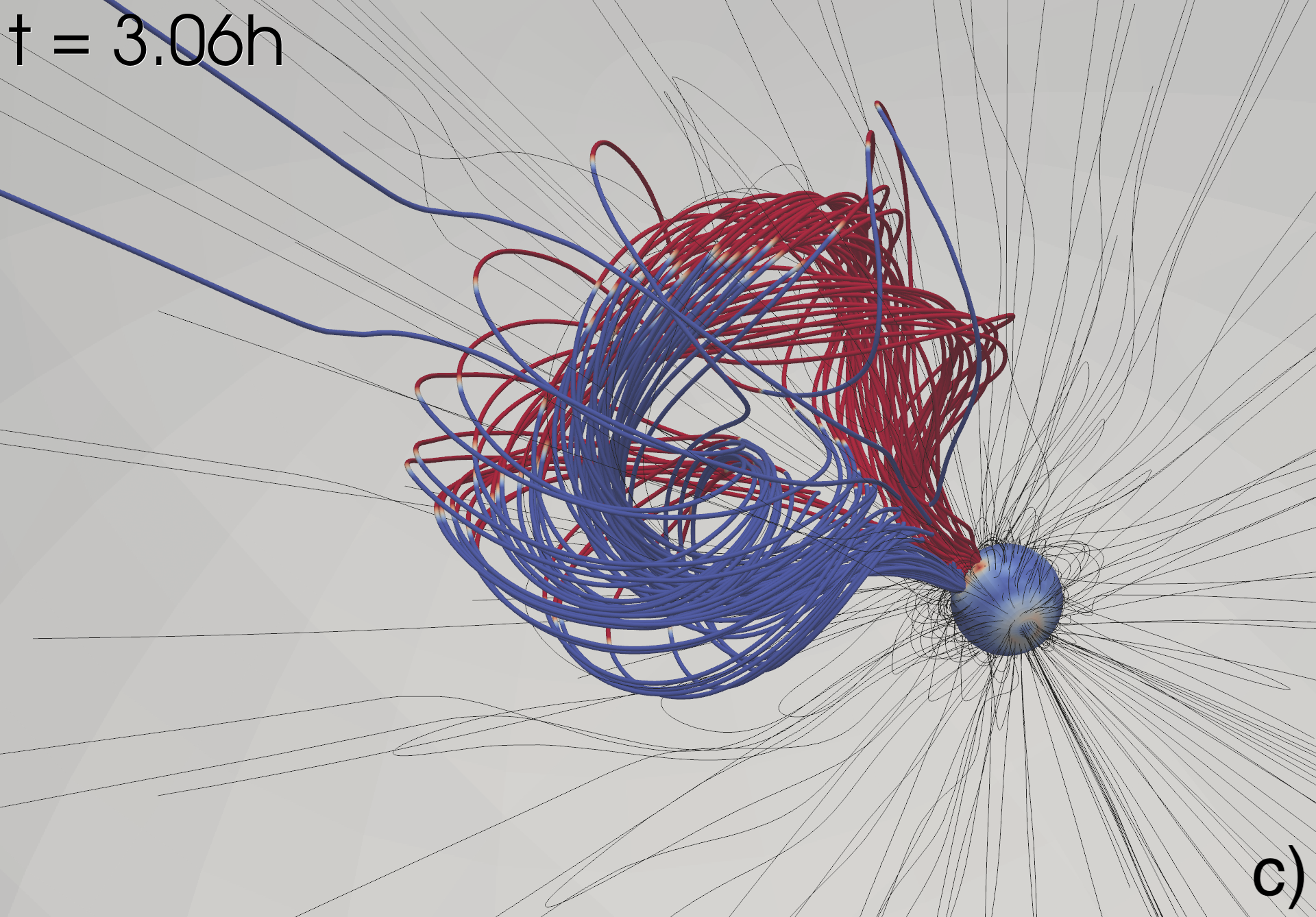}}
    \caption{Visualization of the flux rope CME modeled in COCONUT. The three panels show the evolved flux rope at three different times: at $t = 0.48$\,h (panel a), $t = 1.53$\,h (panel b), and $t = 3.06$\,h (panel c). The sphere of 1\,$R_\odot$ with the mapped radial magnetic field component symbolizes the Sun. The colors of the sphere and flux rope denote the magnetic field's polarity, where red shades describe positive polarity, blue shades describe negative polarity, and white color stands for zero values. The black lines illustrate a sample of the global magnetic field.}
    \label{fig:flux_rope}
\end{figure*}

Figure~\ref{fig:flux_rope} shows the evolving TDFR at three times: $t = 0.48$\,h (panel a), 1.53\,h (panel b), and 3.06\,h (panel c).
The sphere of radius $1\,R_\odot$ symbolizes the Sun. Instead of a color bar with values, we color-coded the polarity (that is, the sign) of the radial magnetic field component $B_r$:
blue shades represent negative polarity ($B_\mathrm{r} <0$), red shades represent positive polarity ($B_\mathrm{r} >0$), and white represents zero values ($B_\mathrm{r} = 0$). The black field lines indicate a sample of the global magnetic field. 
The three panels illustrate the dynamic evolution of the TDFR as the CME propagates through the solar corona. Panel a) highlights the highly twisted magnetic field configuration at an early stage of the CME evolution, with field lines spiraling around the central axis of the TDFR. This twist creates a complex magnetic topology,  characterized by variations both in magnetic field strength and field line curvature. Panels b) and c) show the continued expansion of the flux rope as it propagates radially outward, introducing time-dependent variations in the magnetic field's magnitude and geometry. Moreover, as the TDFR expands, magnetic reconnection (MR) occurs at its footpoints and at the nose of the CME,  altering the magnetic topology.

After the COCONUT simulation, PARADISE takes the obtained coronal configurations to evolve energetic particles as test particles through these backgrounds by solving the focused transport equation (FTE; see, e.g., \citealt{vandenBerg-etal-2020} and references therein). PARADISE solves the FTE grid-free, requiring an interpolation of COCONUT's solar wind parameters and their gradients to the particle's location at each time step. The FTE contains diffusion coefficients for pitch-angle diffusion and CFD to include the effects of turbulence in the solar wind. While initial derivations of the FTE did not include any CFD terms (e.g., \citealt{Zank-2014}), leading to questions about the validity of including a CFD coefficient to the FTE, the picture has changed in the last two decades as several authors (e.g., \citealt{ Zhang-2006, le-Roux-Webb-2007}) derived the FTE with CFD terms using nonlinear theory \citep{Shalchi-2009, Shalchi-2020}. Despite being on a more solid theoretical foundation, it is still highly debated which form of the CFD coefficient to apply, partly due to the significant uncertainties about the turbulence conditions in the heliosphere. For a more detailed discussion about CFD, see \cite{Strauss-etal-2020}.

In the present work, we adopt two different approaches to CFD. The first approach assumes a constant perpendicular MFP $\lambda_\perp$ and has been employed multiple times in the literature due to its simplicity, for instance, by \cite{Wijsen-etal-2019b} or \cite{Husidic-etal-2024}. In the second approach, we choose 
$\lambda_\perp = (\pi/12)\,\alpha\,(r_\mathrm{L}/r_\mathrm{L,0})\,\lambda_\parallel$, 
where $\lambda_\parallel$ is the parallel MFP, $r_\mathrm{L}$ is the maximum Larmor radius (see Eq.~\ref{eq:larmor_radius}), and $r_\mathrm{L,0}$ is a reference Larmor radius.  This approach has been adopted previously by, for instance, \cite{Droege-etal-2010}, \cite{Wijsen-etal-2019a}, \cite{Laitinen-Dala-2021} and \cite{Niemela-etaL-2024}. 
For all simulations in Sec.~\ref{sec:results}, we choose a constant parallel MFP $\lambda_\parallel = 21.5\,R_\odot$ (0.1\,au), corresponding to the radius of the simulation domain\footnote{At the latest time steps, some twisted field lines of the flux rope can have lengths of up to $10 \times \lambda_\parallel$.}. In addition, for the simulations in Sec.~\ref{subsec:larmor}, 
we assume a reference Larmor radius (see Sec.~\ref{app:paradise}) for a 1\,MeV proton in a magnetic field of strength $B_0 = 30$\,nT, corresponding to a typical field strength value at the outer boundary 
(for comparison, $B$ at the inner boundary point is $\sim 5\times 10^4$\,nT). This way, only the parameter $\alpha$ is varied to determine the effect of different levels of CFD on the particle distributions.

In all simulations, we uniformly inject particles within the radial range of $1.49,R_\odot$ to $1.50,R_\odot$, specifically in regions where the magnetic field strength exceeds $9\times 10^4$~nT, and the magnetic polarity is inward. This effectively places the initial particle distribution near the central axis of the flux rope's leg with negative polarity, close to the inner boundary of the simulation domain. 
We inject an isotropic (in pitch-angle) monoenergetic distribution of 100~keV protons, with all particles being introduced at $t = 0.46$~h into the simulation (where $t = 0$ corresponds to the flux rope insertion). By this time, the nose of the erupting flux rope has reached a distance of approximately $3\,R_\odot$. Since the FTE is solved in a stochastic manner, we injected a total of 3.6 million pseudo-particles to guarantee adequate statistics. Finally, we use absorbing conditions at PARADISE's inner and outer boundaries.

\section{Results}\label{sec:results}

\subsection{Simulation without CFD}\label{subsec:no_CFD}

\begin{figure*}[t]
    \centering
    \subfigure{\includegraphics[width=0.40\textwidth]{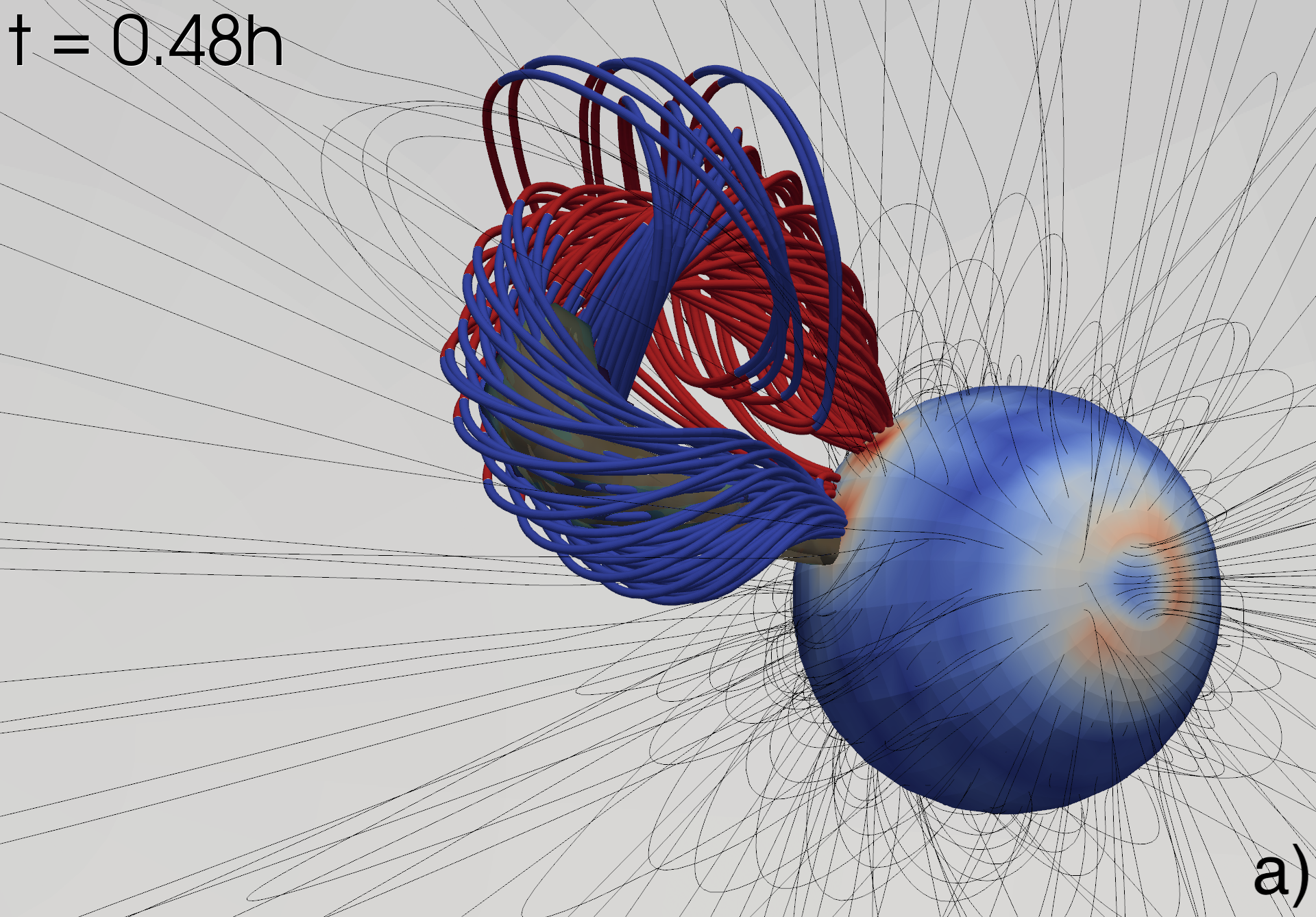}} \quad
    \subfigure{\includegraphics[width=0.40\textwidth]{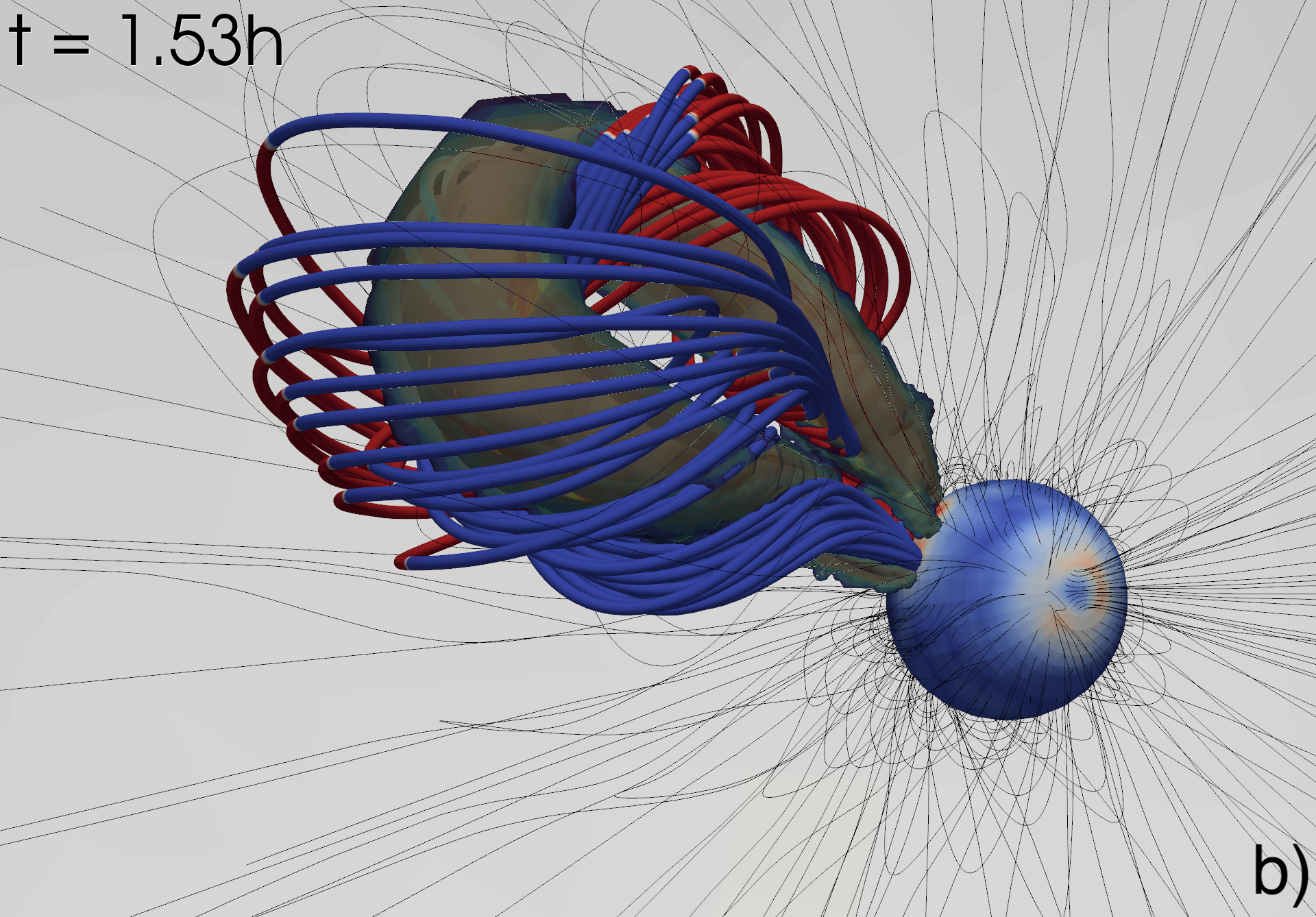}} \\
    \subfigure{\includegraphics[width=0.40\textwidth]{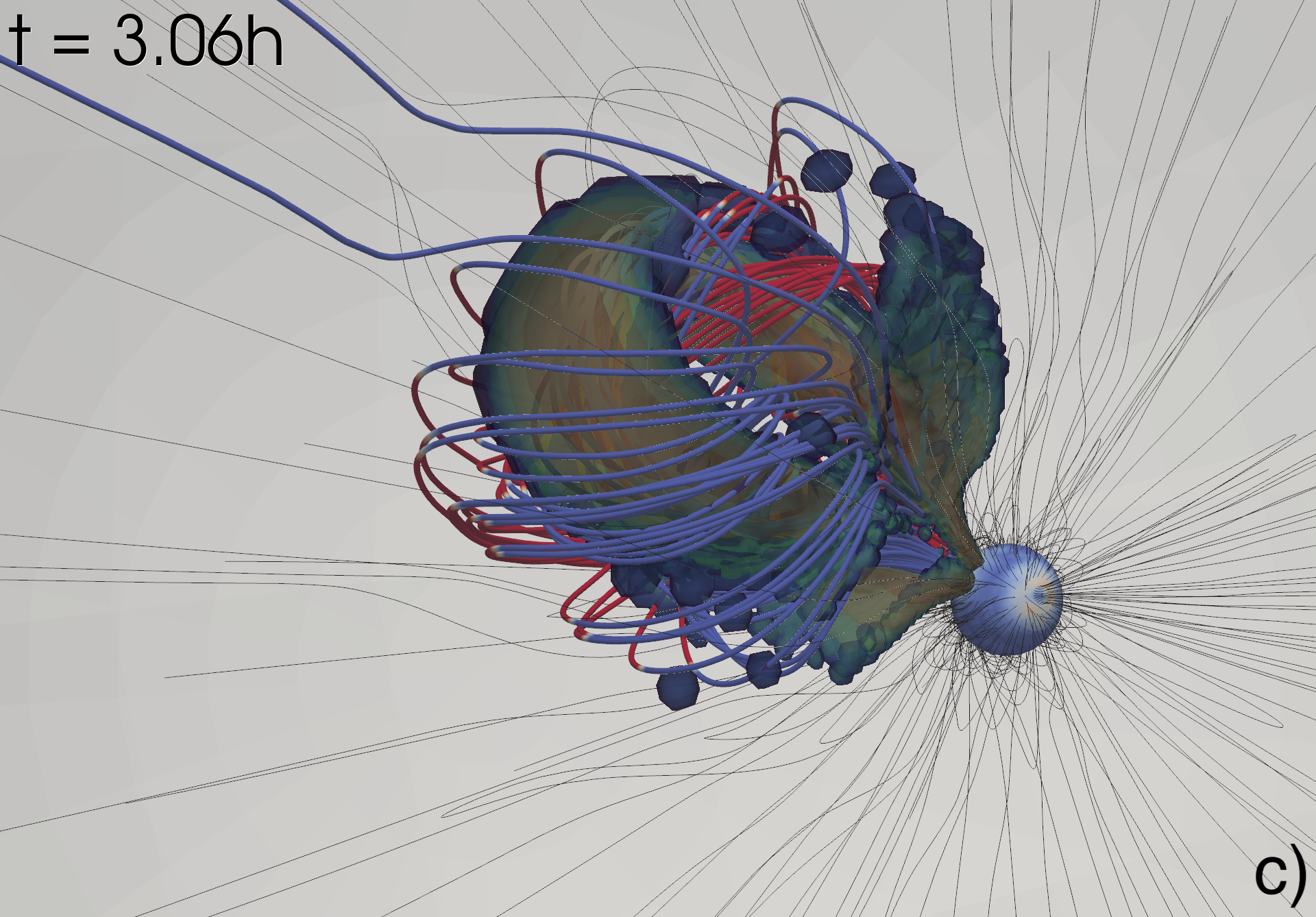}} \quad
    \subfigure{\includegraphics[width=0.40\textwidth]{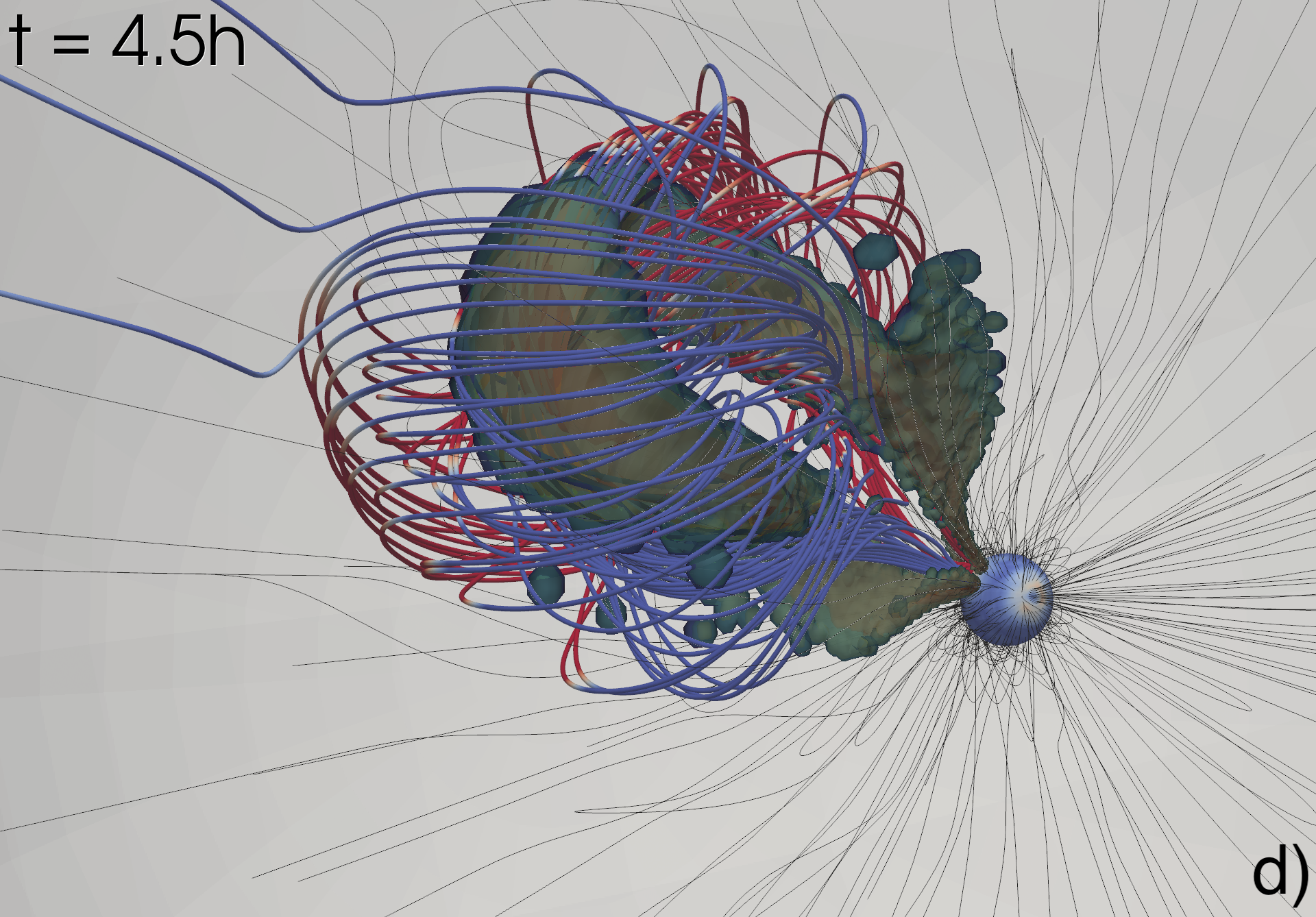}} \\
    \begin{minipage}[b]{0.8\textwidth}
        \includegraphics[width=\textwidth]{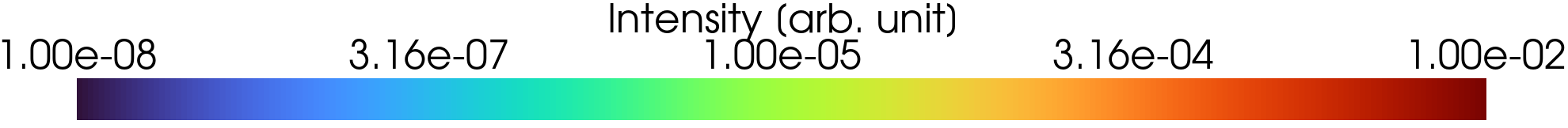} 
    \end{minipage}
    \caption{Time-lapse of particle propagation in the flux rope without CFD. Panels a) through d) show particle intensity contours at $t = $ 0.48\,h, 1.53\,h, 3.06\,h, and 4.5\,h, respectively, plotted together with the TDFR. The sphere of 1 solar radius with the mapped radial magnetic field component symbolizes the Sun. }
    \label{fig:no_cfd}
\end{figure*}

Figure~\ref{fig:no_cfd} contains a time-lapse of a particle transport simulation without any CFD (i.e., $\lambda_\perp = 0$). The four panels show 3D plots of the evolving flux rope together with contours of the particle intensities at four different times: 0.48\,h, 1.53\,h, 3.06\,h, and 4.5\,h in panels a) to d), respectively. Panel a) of Fig.~\ref{fig:no_cfd} indicates that after their injection at the bottom of the flux rope leg, the particles have moved up along the interior flux rope field lines, which follow the central, arc-shaped structure of the TDFR, and remain confined to these field lines. In panel b), it can be noticed that the particles reached the opposite footpoint of the flux rope. While some particles are magnetically reflected and move back towards the original leg of the flux rope, other particles fall back to the Sun and are removed from the simulation. In panel c), about 2.5\,h after injection, some particles at the base of the TDFR gain access to its exterior magnetic field lines, which wrap around the interior ones, and start to move along them. Finally, in panel d), it is evident that some particles propagate along the exterior field lines of the TDFR. This behavior occurs due to MR between the interior and the exterior magnetic field lines of the TDFR near footpoints and is influenced by numerical diffusion. Any ideal MHD simulation will include numerical diffusion and thus induce artificial MR. However, the MR occurs in regions where we expect real MR to occur. The highest particle intensities remain concentrated along the central axis of the flux rope. Furthermore, open magnetic field lines are visible in panels c) and d); however, without the CFD mechanism, the protons with an initial energy of 100\,keV do not gain access to those or the global magnetic field lines in our simulation and stay confined to the TDFR. MR also occurs between the TDFR field lines and the global magnetic field; hence, we expect particles to gain access to those field lines as well eventually.

\subsection{Constant Perpendicular MFP Length}\label{subsec:const_mfp_perp}

\begin{figure*}[t]
    \centering
    \subfigure{\includegraphics[width=0.30\textwidth]{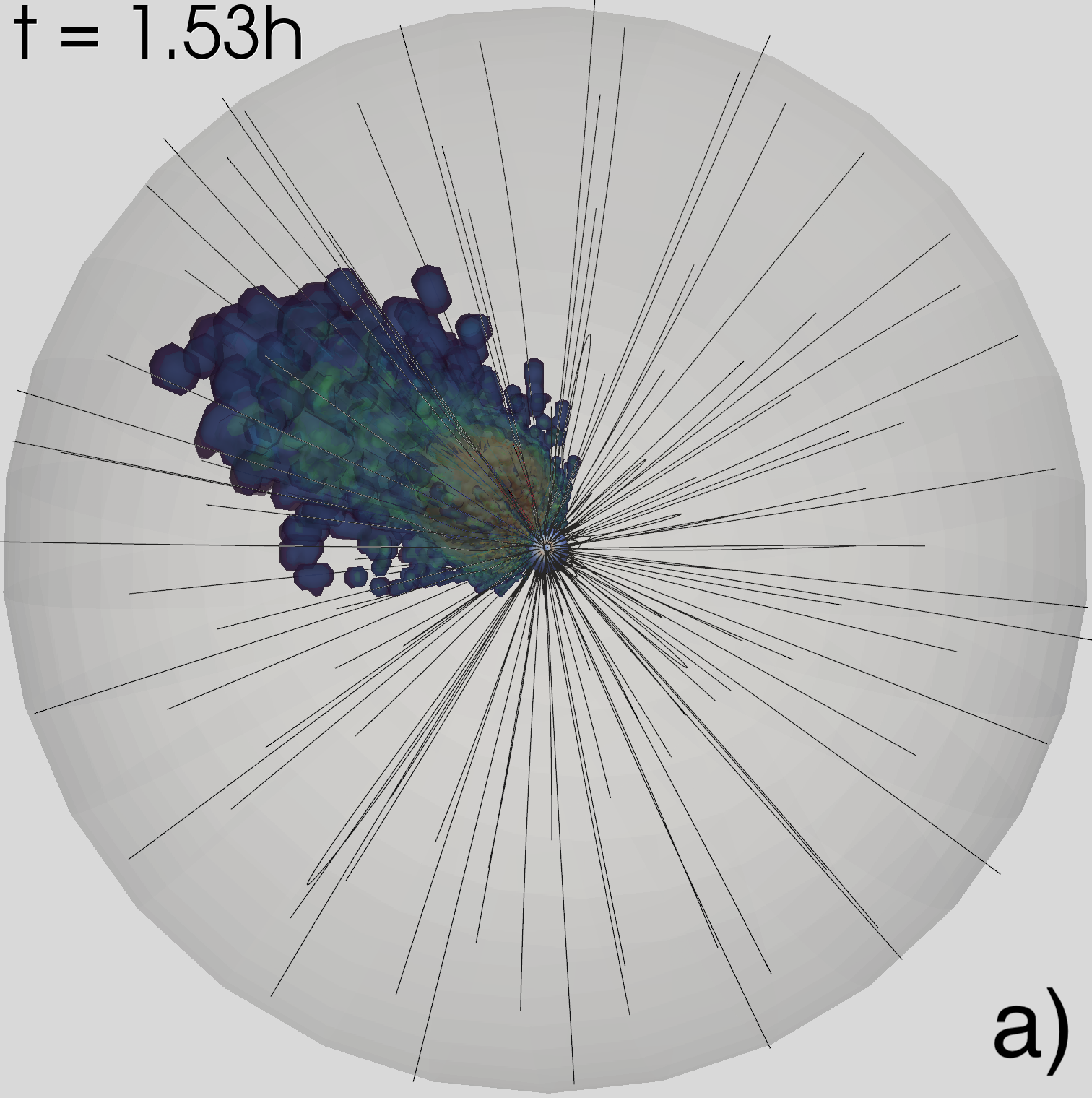}} \quad
    \subfigure{\includegraphics[width=0.30\textwidth]{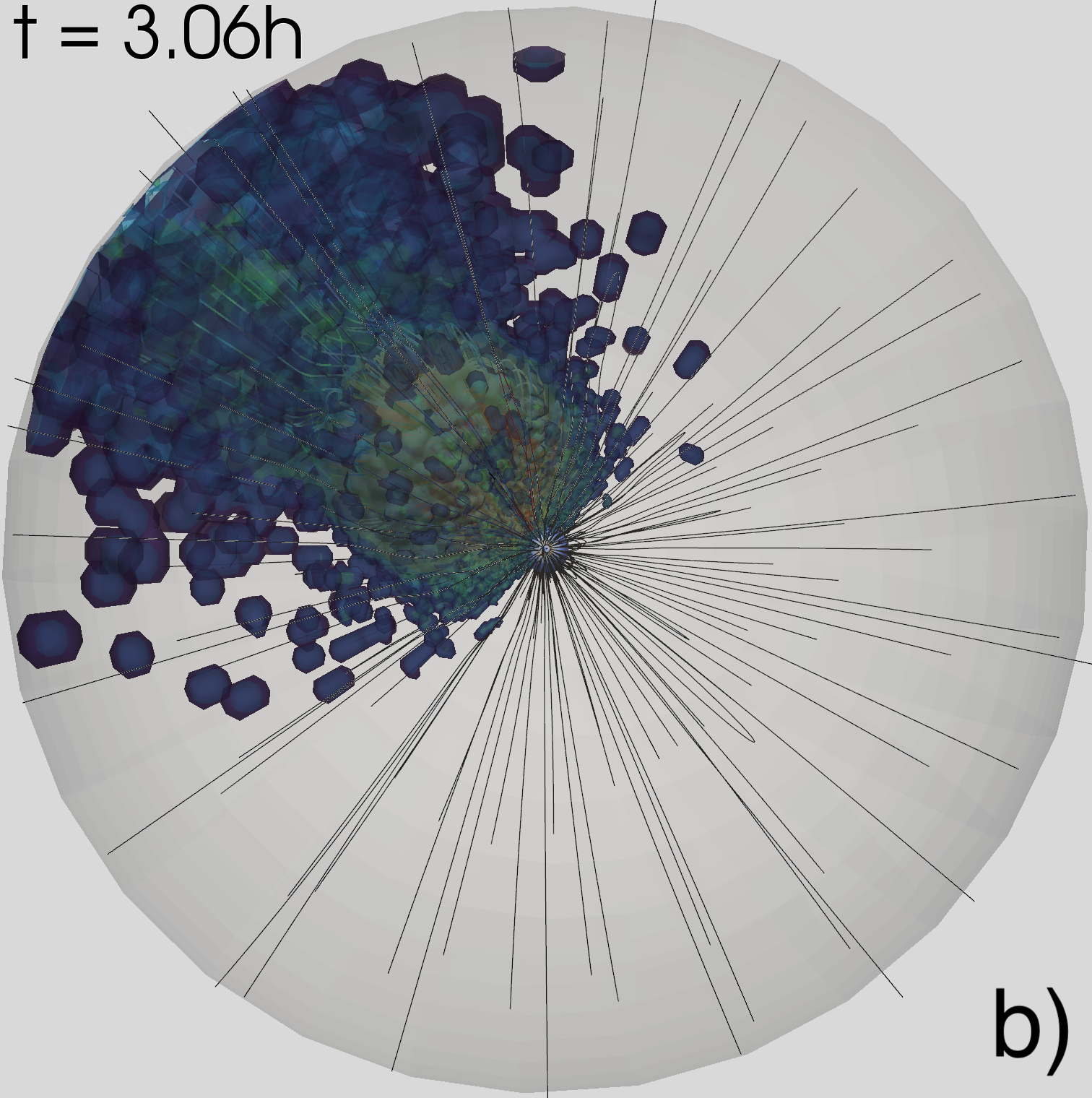}} \quad
    \subfigure{\includegraphics[width=0.30\textwidth]{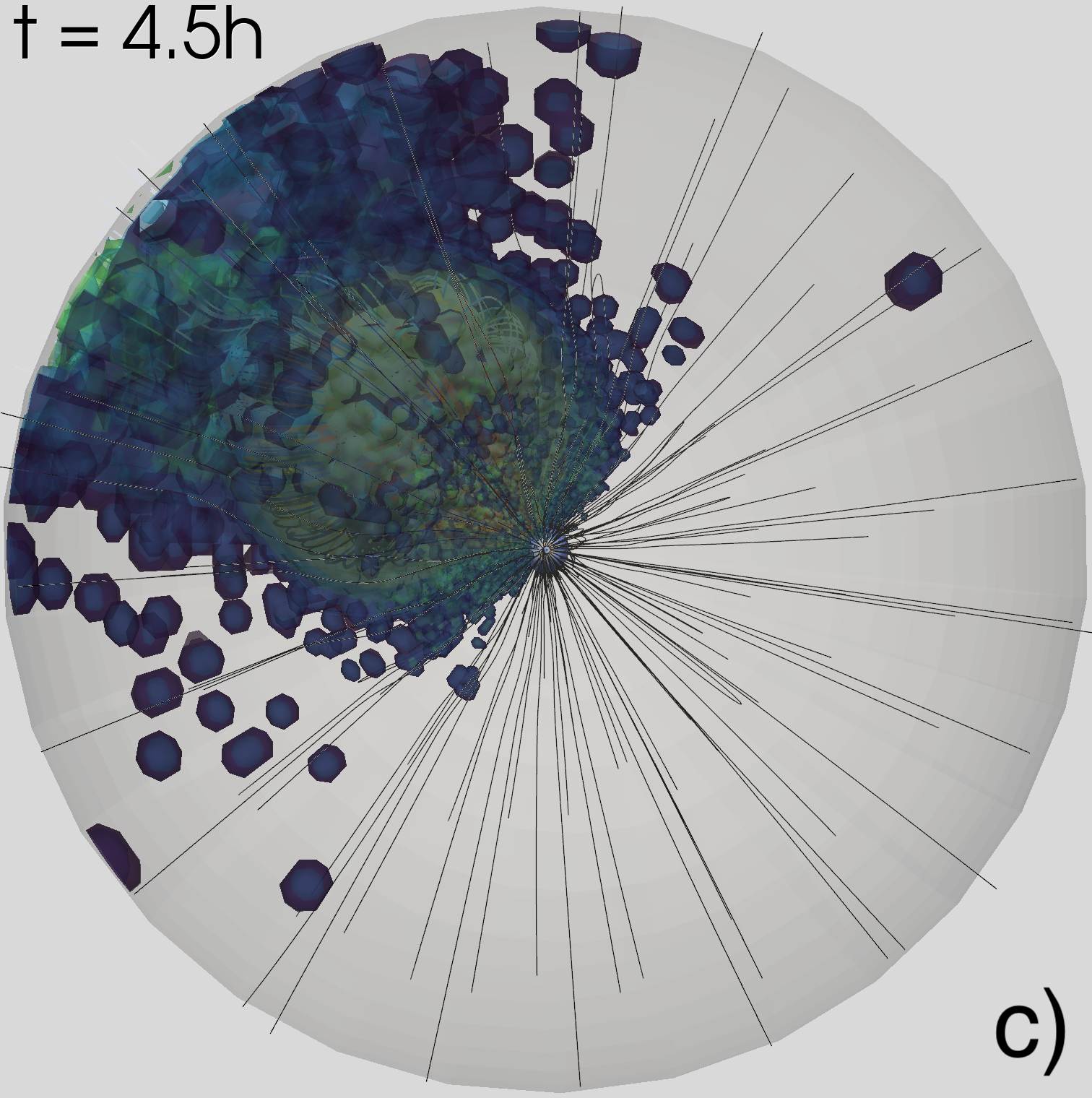}} \\
    \subfigure{\includegraphics[width=0.30\textwidth]{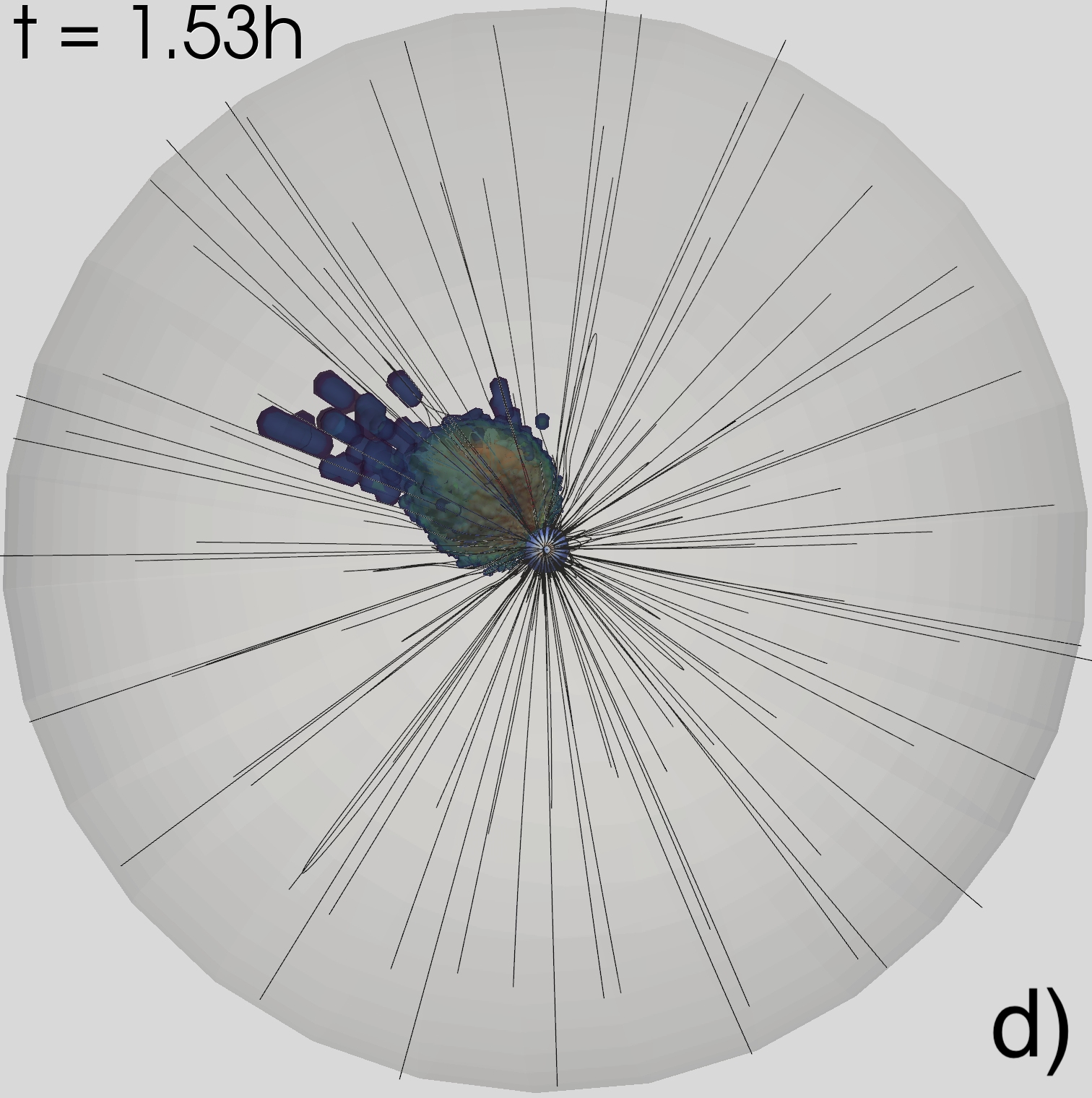}} \quad
    \subfigure{\includegraphics[width=0.30\textwidth]{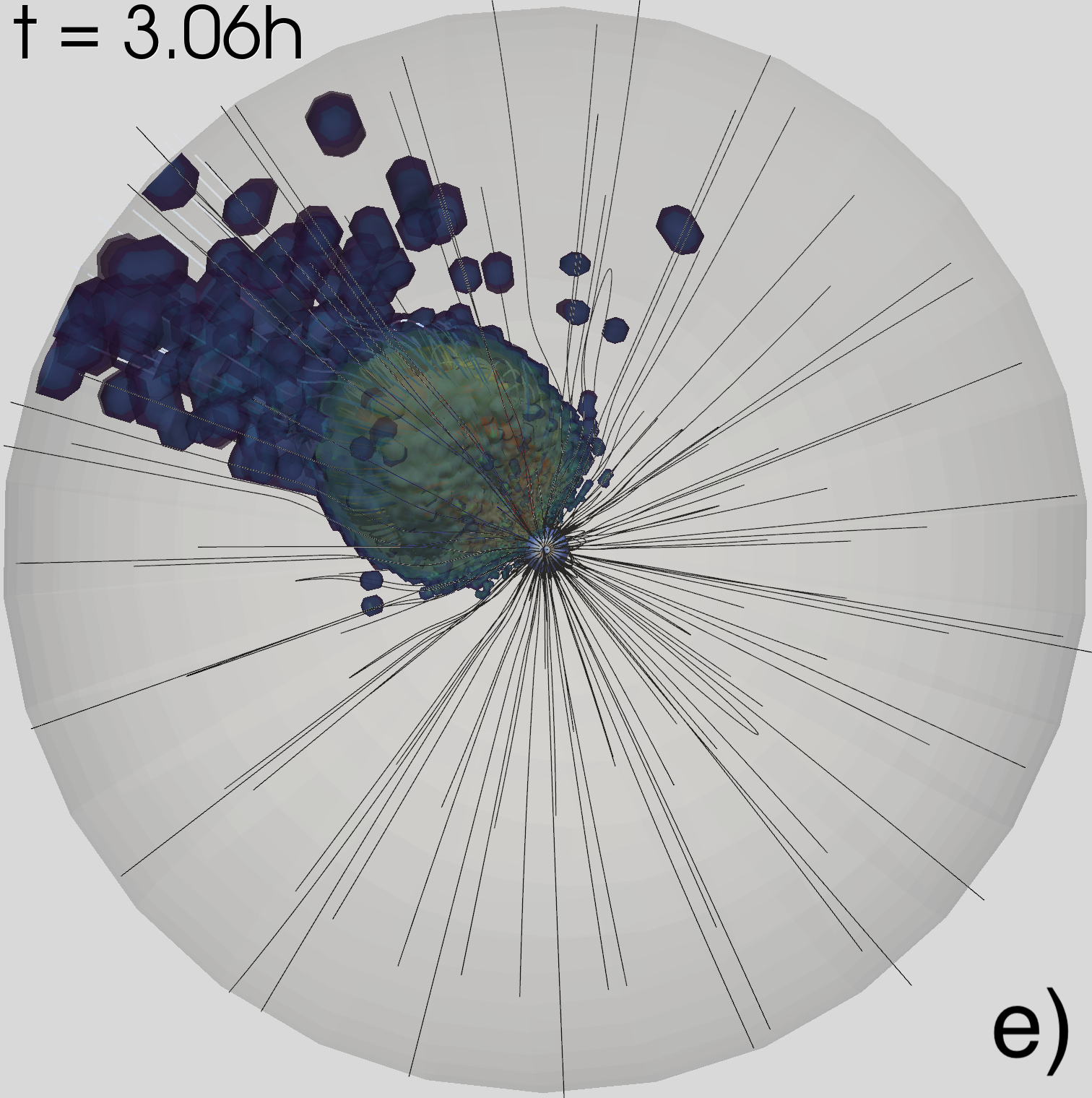}} \quad
    \subfigure{\includegraphics[width=0.30\textwidth]{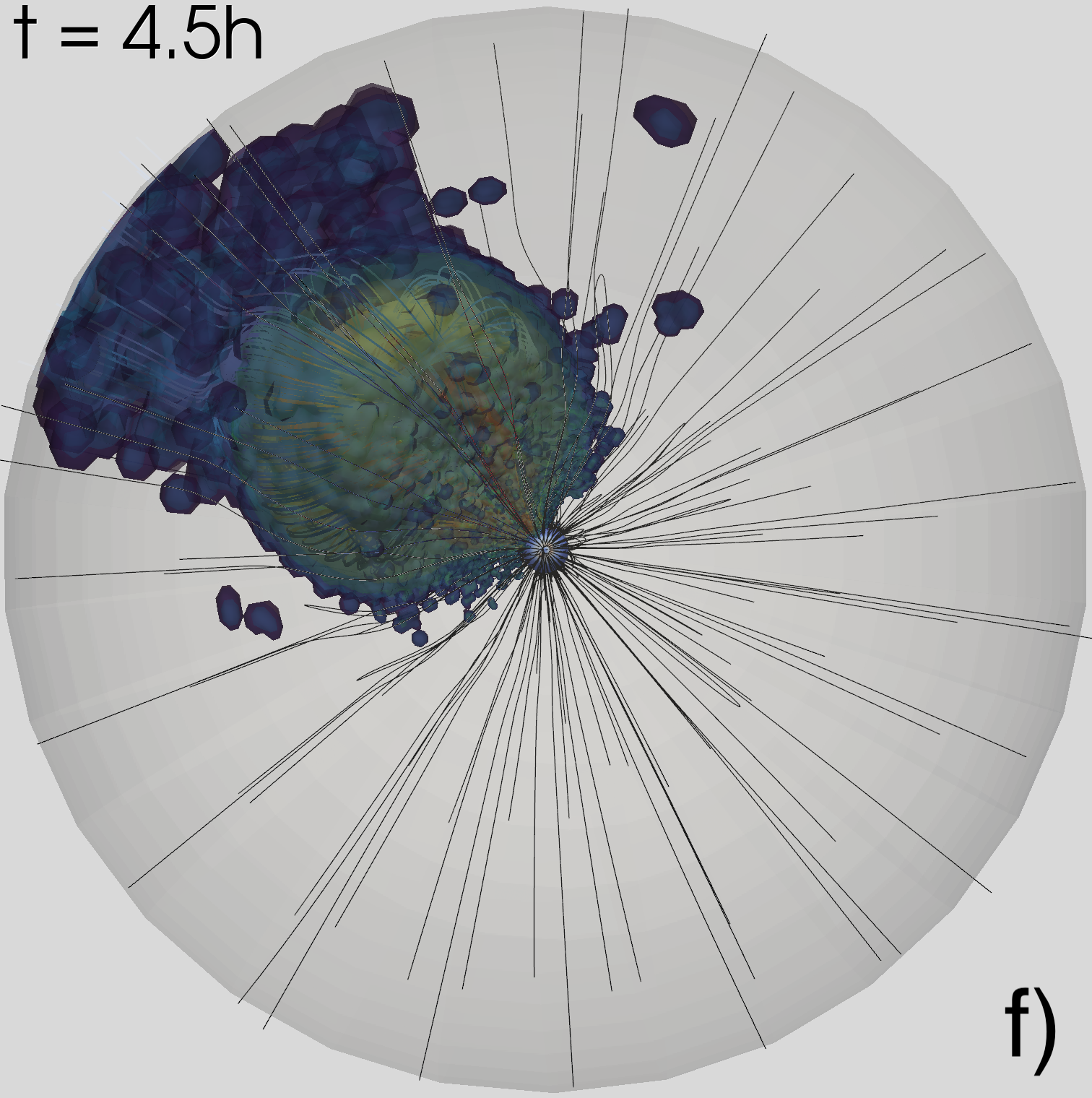}}\\
    \subfigure{\includegraphics[width=0.30\textwidth]{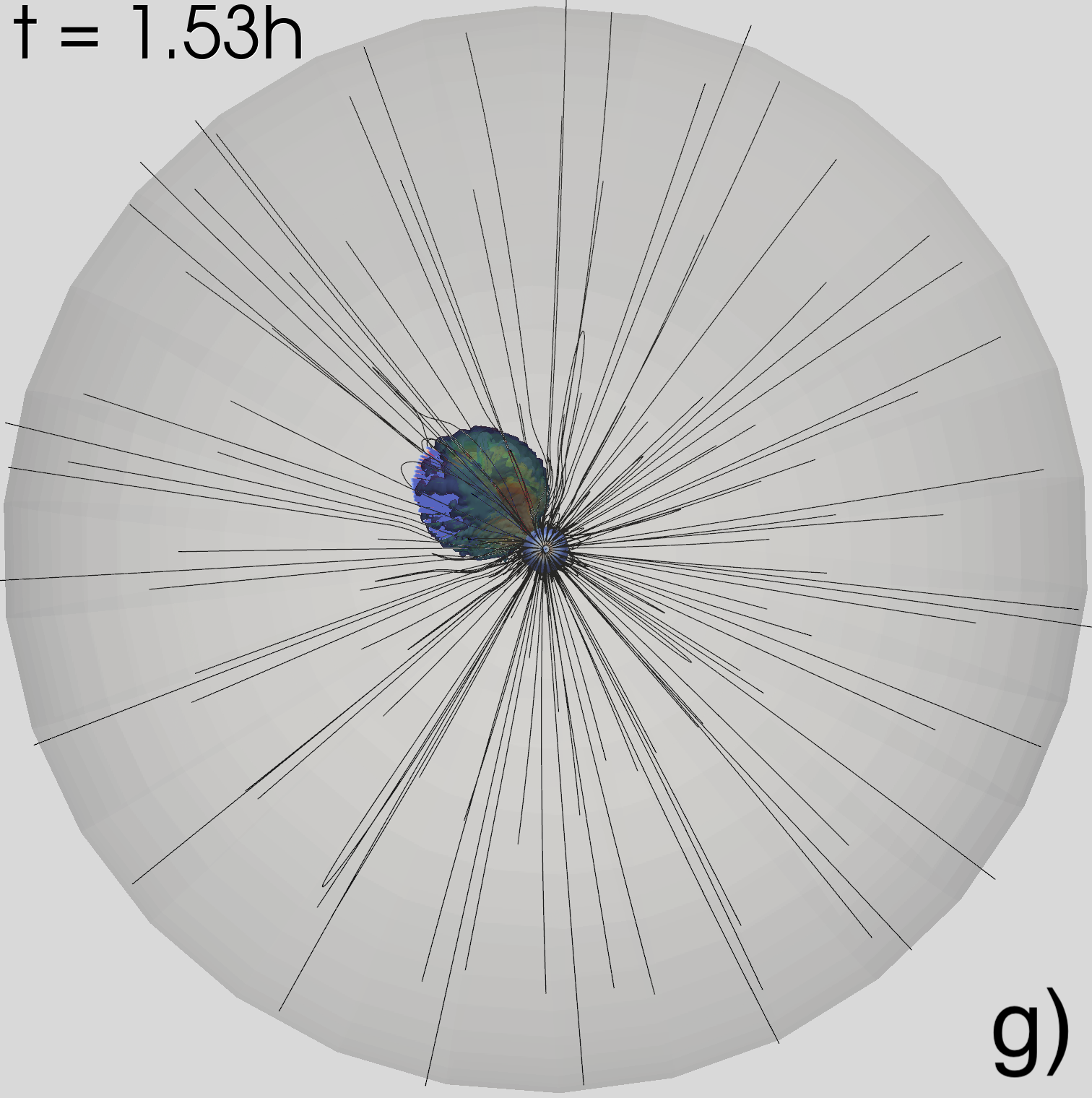}} \quad
    \subfigure{\includegraphics[width=0.30\textwidth]{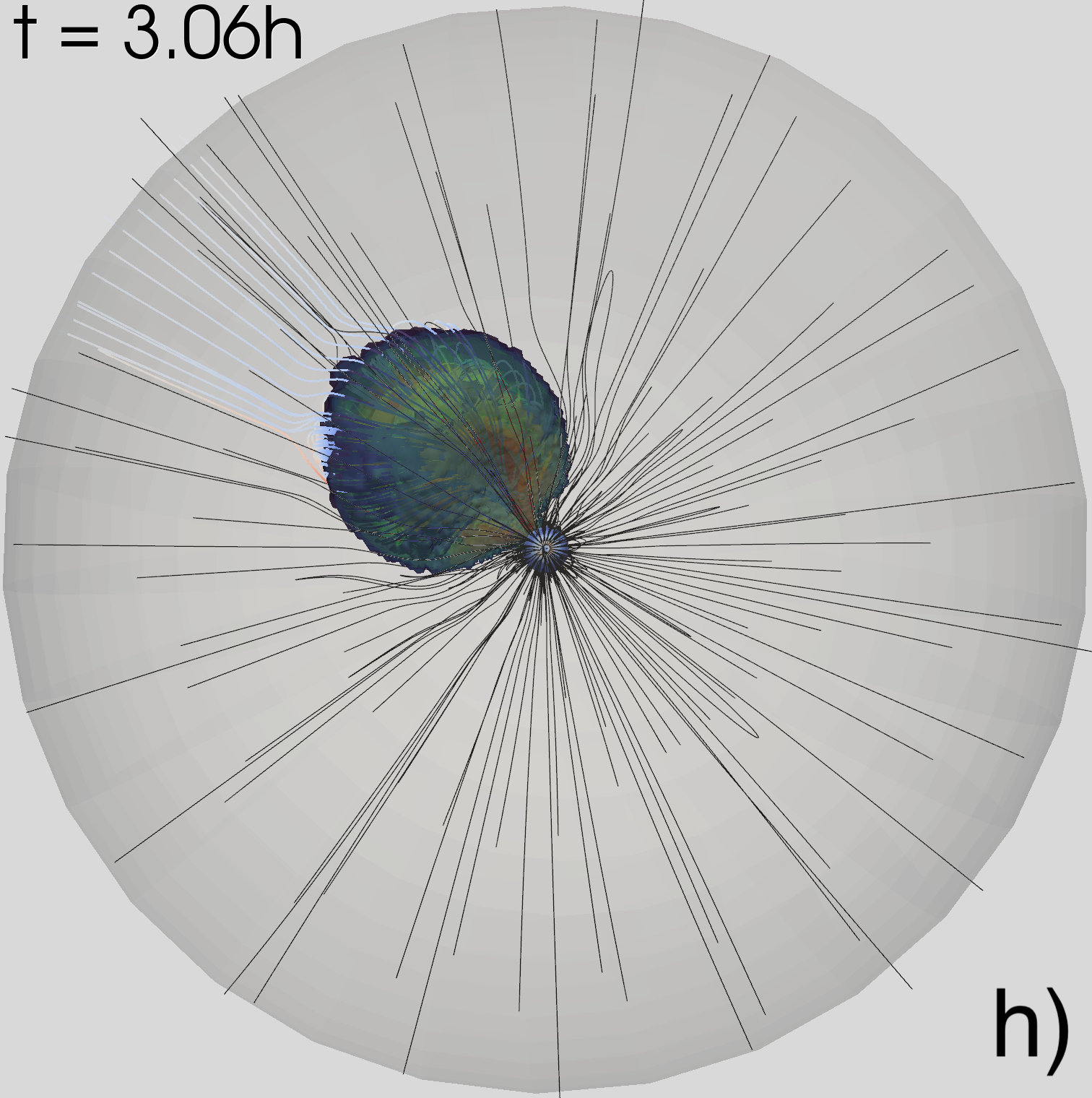}} \quad
    \subfigure{\includegraphics[width=0.30\textwidth]{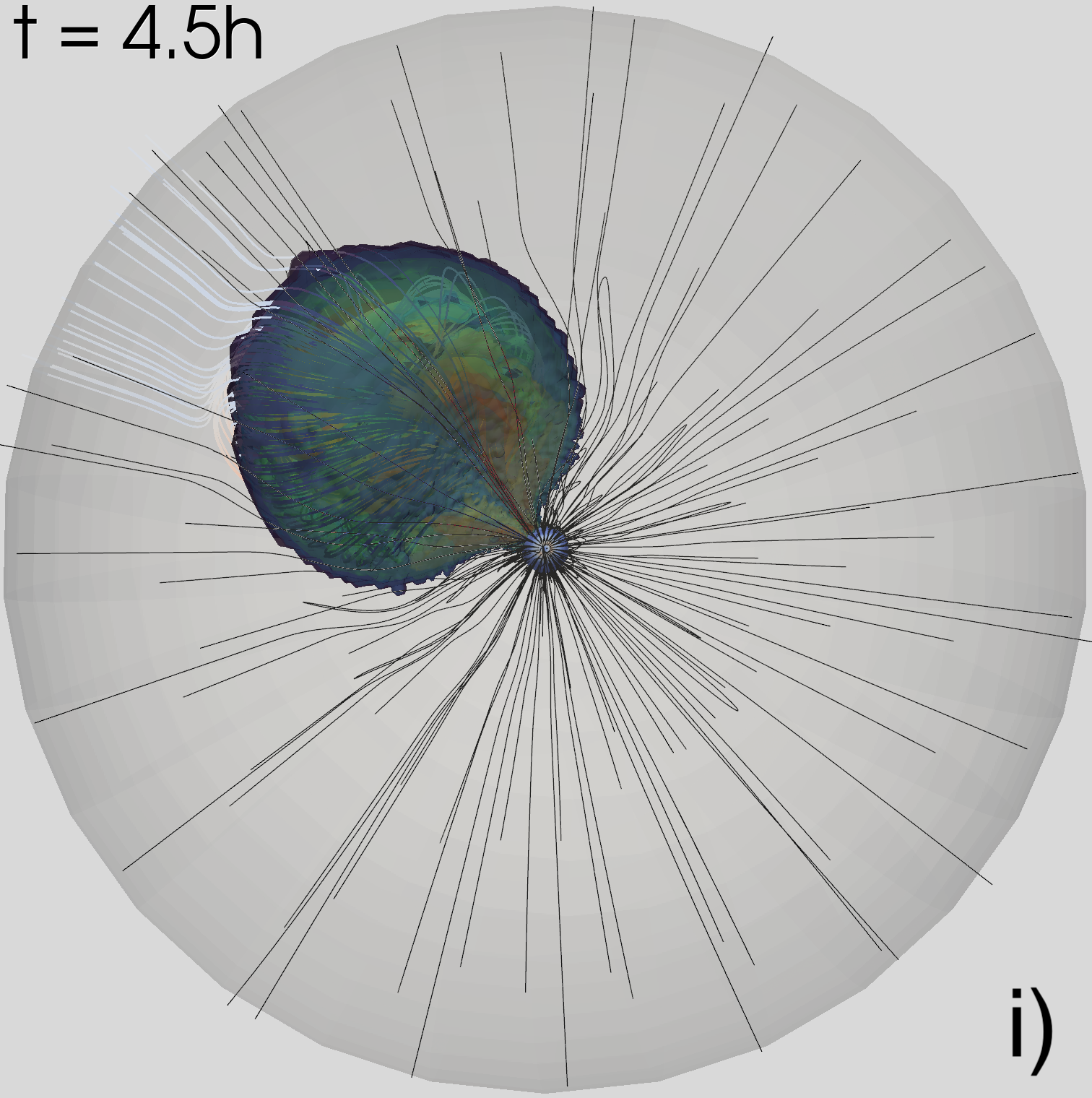}}
    \caption{Time-lapse of particle propagation in the flux rope with CFD using a constant perpendicular MFP $\lambda_\perp$. The top panels (a through c) show the results for $\lambda_\perp = 2.150\times 10^{-2}\,R_\odot$, the middle panels (d through f) for $\lambda_\perp = 1.075\times 10^{-2}\,R_\odot$, and the bottom panels (g through i) for $\lambda_\perp = 2.150\times 10^{-3}\,R_\odot$. The three columns correspond to the times 1.53\,h, 3.06\,h, and 4.5\,h, respectively. The same color bar as in Fig.~\ref{fig:no_cfd} is used for the intensities.}
    \label{fig:const_mfp_perp}
\end{figure*}

Figure~\ref{fig:const_mfp_perp} displays particle transport simulations with CFD using a constant perpendicular MFP length $\lambda_\perp$. The panels are ordered such that the three columns correspond from left to right to the times at 1.53\,h, 3.06\,h, and 4.5\,h, respectively, while the three rows correspond from top to bottom to the constant $\lambda_\perp$-values $2.150 \times 10^{-2}\,R_\odot$, $ 1.075 \times 10^{-2}\,R_\odot$, and $ 2.150 \times 10^{-3}\,R_\odot$, respectively. 

In the top row with $\lambda_\perp = 2.150 \times 10^{-2}\,R_\odot$ (or $10^{-4}$\,au), we used a value that is commonly employed in heliospheric simulations (e.g., \citealt{Wijsen-etal-2019b, Husidic-etal-2024}). However, despite the small ratio $\lambda_\perp / \lambda_\parallel = 10^{-3}$, we see that CFD has a significant effect. Comparing panel a) of Fig.~\ref{fig:const_mfp_perp} to panel a) of Fig.~\ref{fig:no_cfd}, we see that the particles have not only diffused along the exterior field lines of the flux rope, but they are escaping the CME site, mainly in the propagation direction of the CME. Panels b) and c) display this effect intensified at later steps, indicating that particles also diffuse longitudinally. Once particles have diffused onto the open field lines ahead of the CME, they can easily escape the simulation domain through parallel transport aided by magnetic focusing.

The middle row with $\lambda_\perp = 1.075 \times 10^{-2}\,R_\odot$ and $\lambda_\perp / \lambda_\parallel = 5\times 10^{-4}$ shows considerably less CFD. In panel d), particles have already accessed the exterior field lines of the TDFR but remain largely confined to it, with some particles escaping the CME site in the propagation direction of the CME. This behavior aligns with our expectations, as the propagation direction is where significant MR and, thus, the opening of flux rope field lines is anticipated. 
In panels e) and f), more particles have escaped the TDFR, and compared to the case in the top row, the particles spread much less longitudinally.

In the last row of Fig.~\ref{fig:const_mfp_perp}, where  $\lambda_\perp = 2.150 \times 10^{-3}\,R_\odot \approx 1500$~km and $\lambda_\perp / \lambda_\parallel = 10^{-5}$, as expected, the least amount of CFD is observed. In all three panels, it can be observed that the particles have enveloped the CME but remain primarily confined to it. However, in Fig.~\ref{fig:I_integrated} in Sec.~\ref{subsec:larmor}, it is shown that even in this case, some particles manage to escape the flux rope along the opening magnetic field lines at the nose of the flux rope. In general, the results indicate that varying $\lambda_\perp$ within one order already drastically changes the effect of CFD on particle transport.

\subsection{Perpendicular MFP Length Dependent on the Larmor Radius}\label{subsec:larmor}

\begin{figure*}[t]
    \centering
    \subfigure{\includegraphics[width=0.30\textwidth]{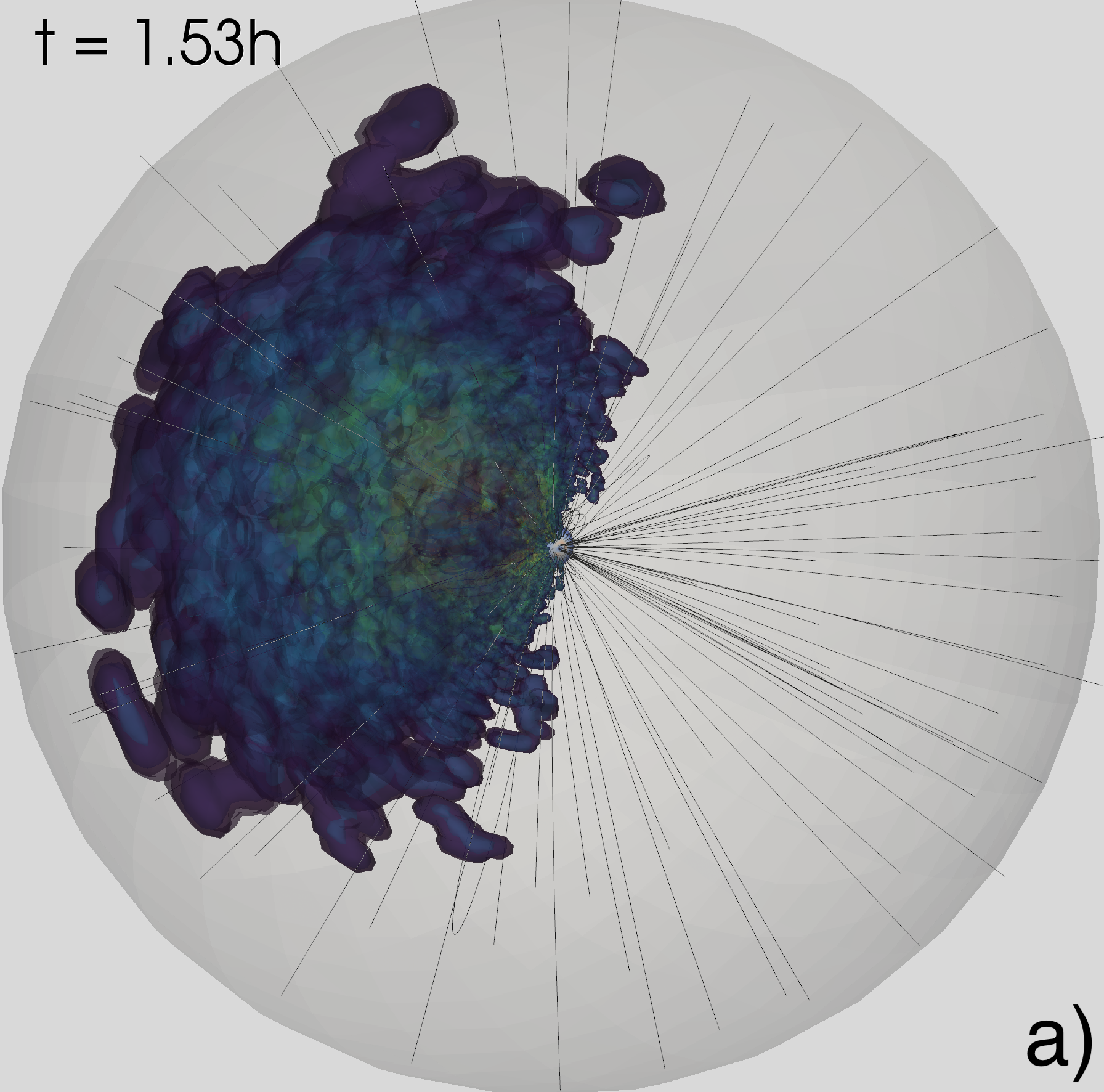}} \quad
    \subfigure{\includegraphics[width=0.30\textwidth]{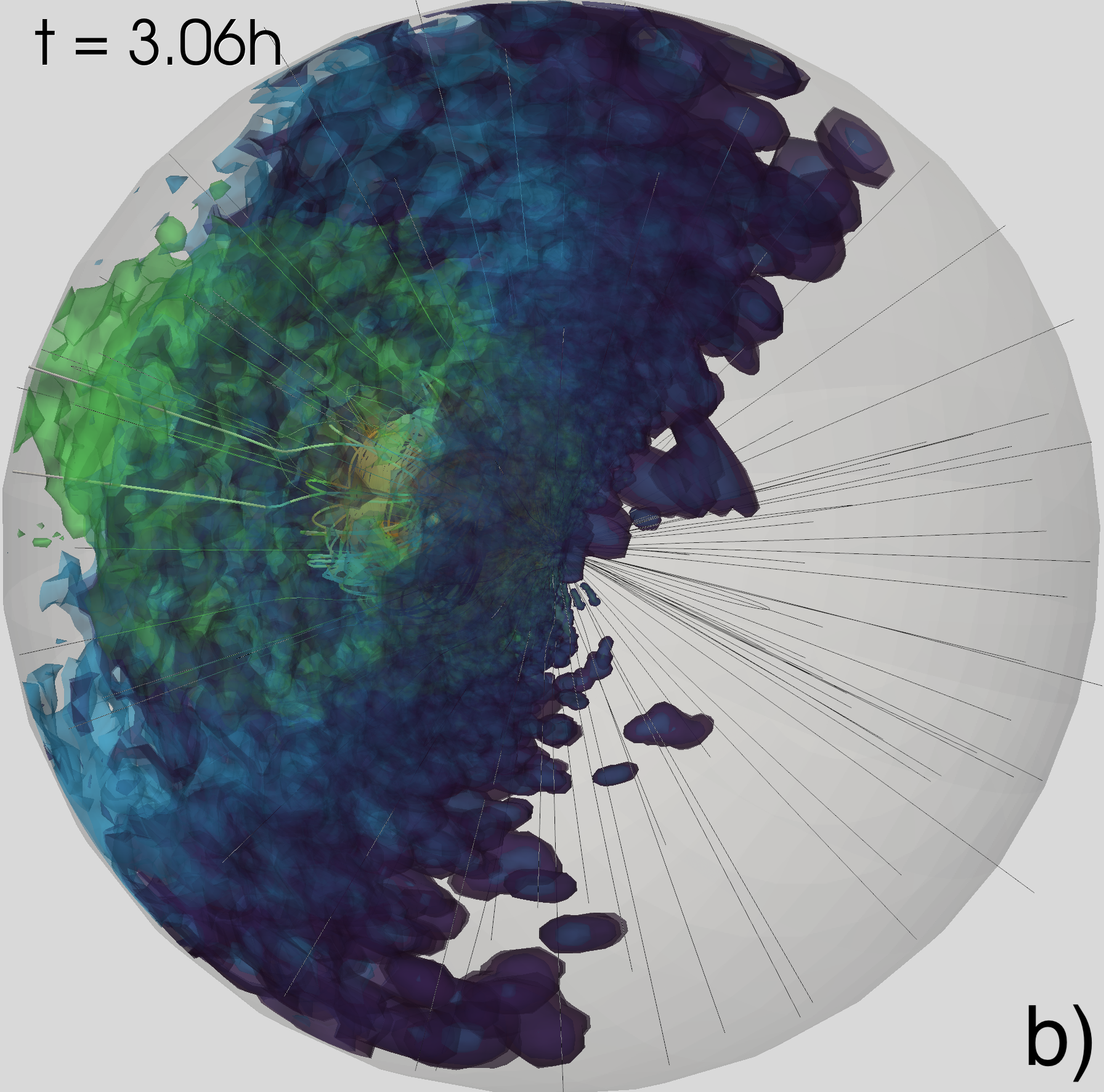}} \quad
    \subfigure{\includegraphics[width=0.30\textwidth]{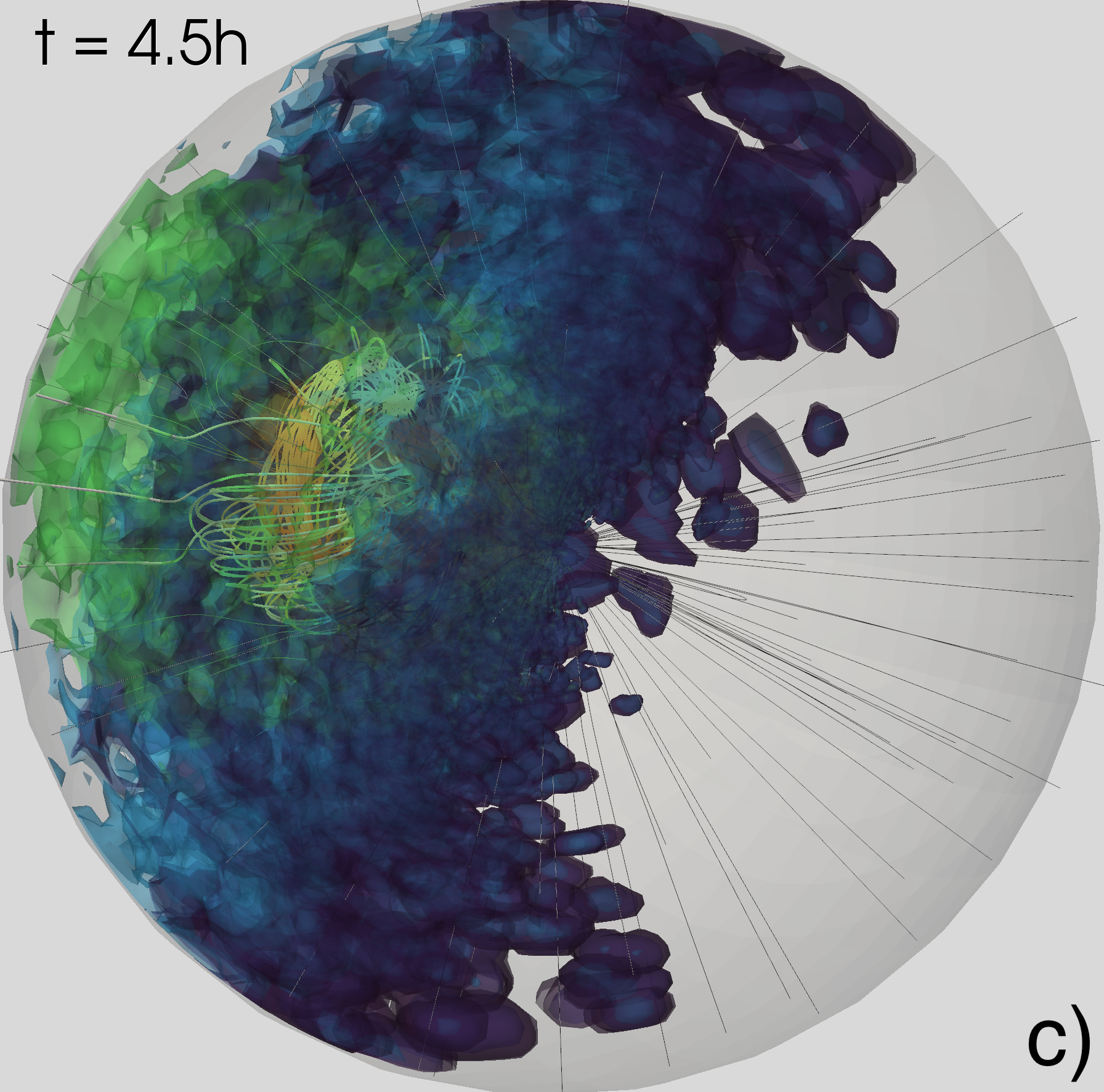}} \\
    \subfigure{\includegraphics[width=0.30\textwidth]{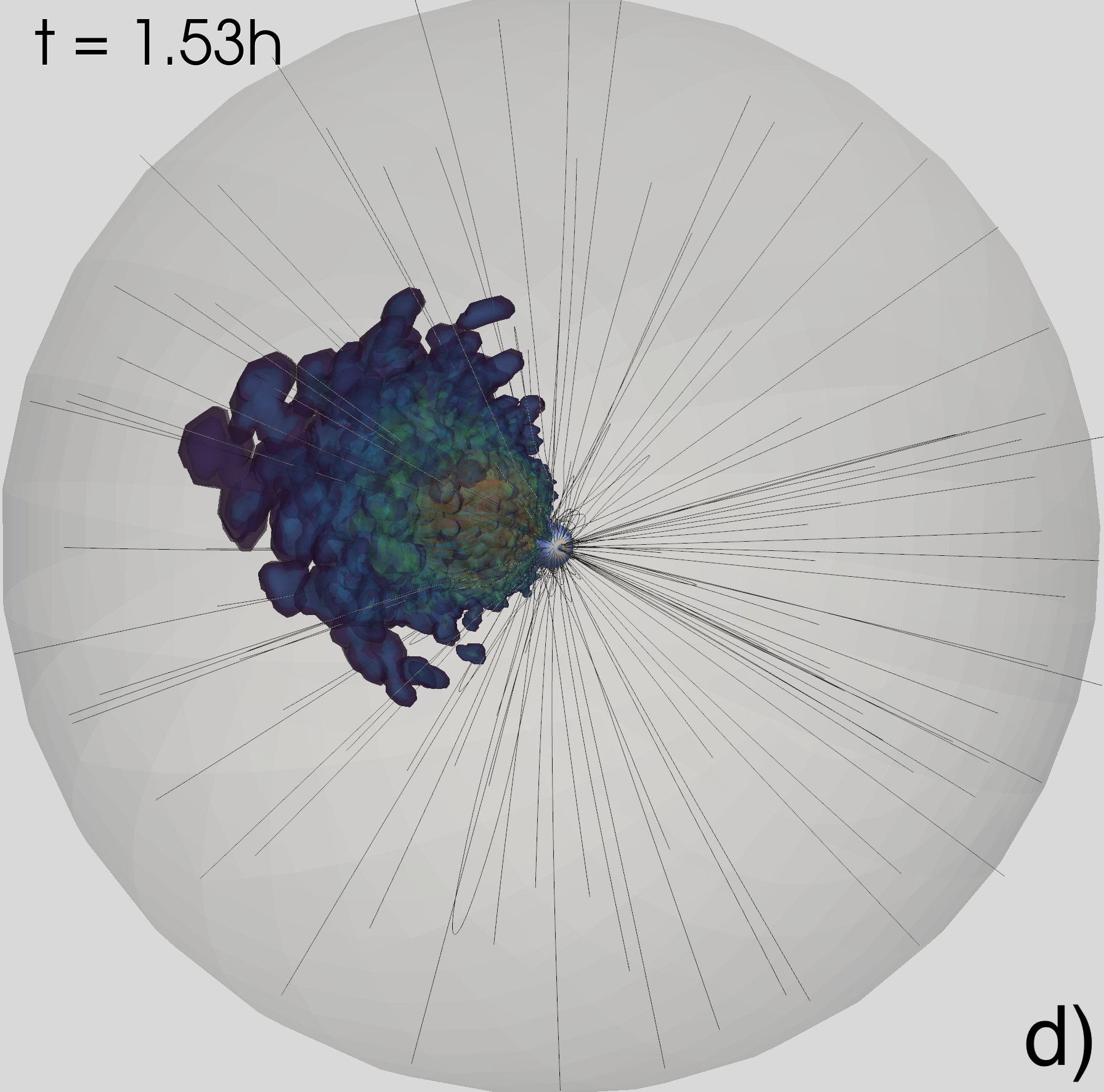}} \quad
    \subfigure{\includegraphics[width=0.30\textwidth]{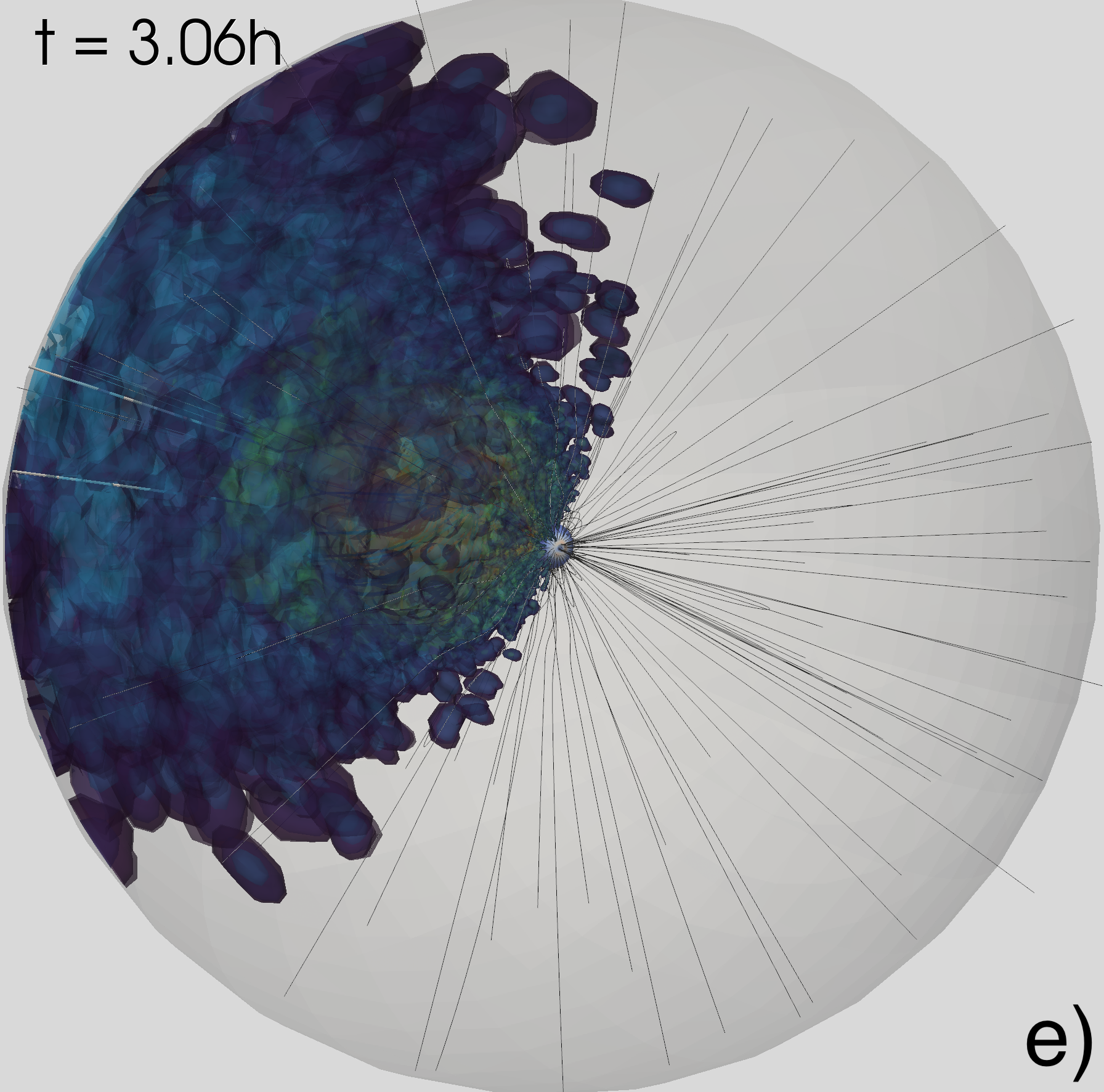}} \quad
    \subfigure{\includegraphics[width=0.30\textwidth]{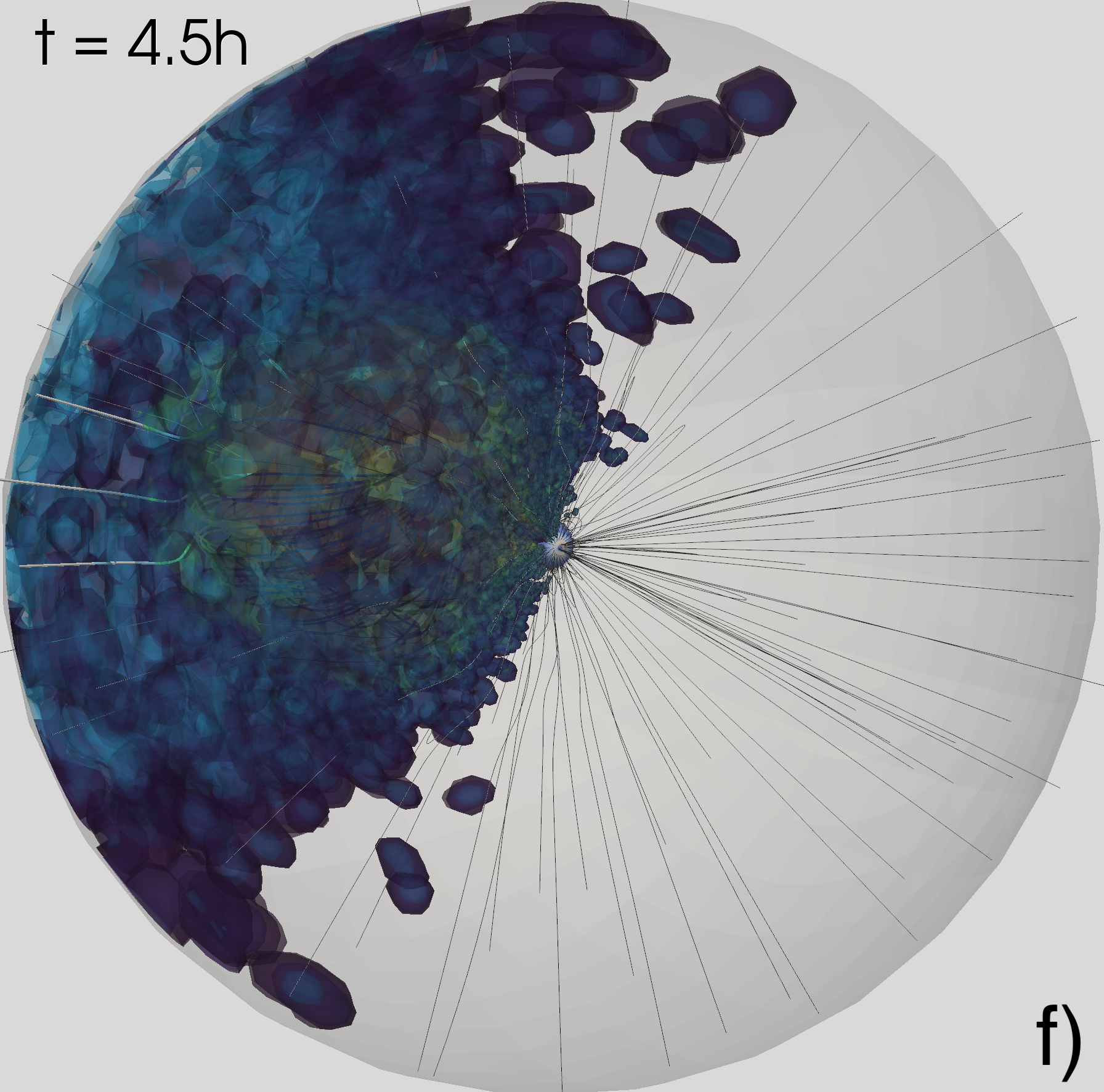}}\\
    \subfigure{\includegraphics[width=0.30\textwidth]{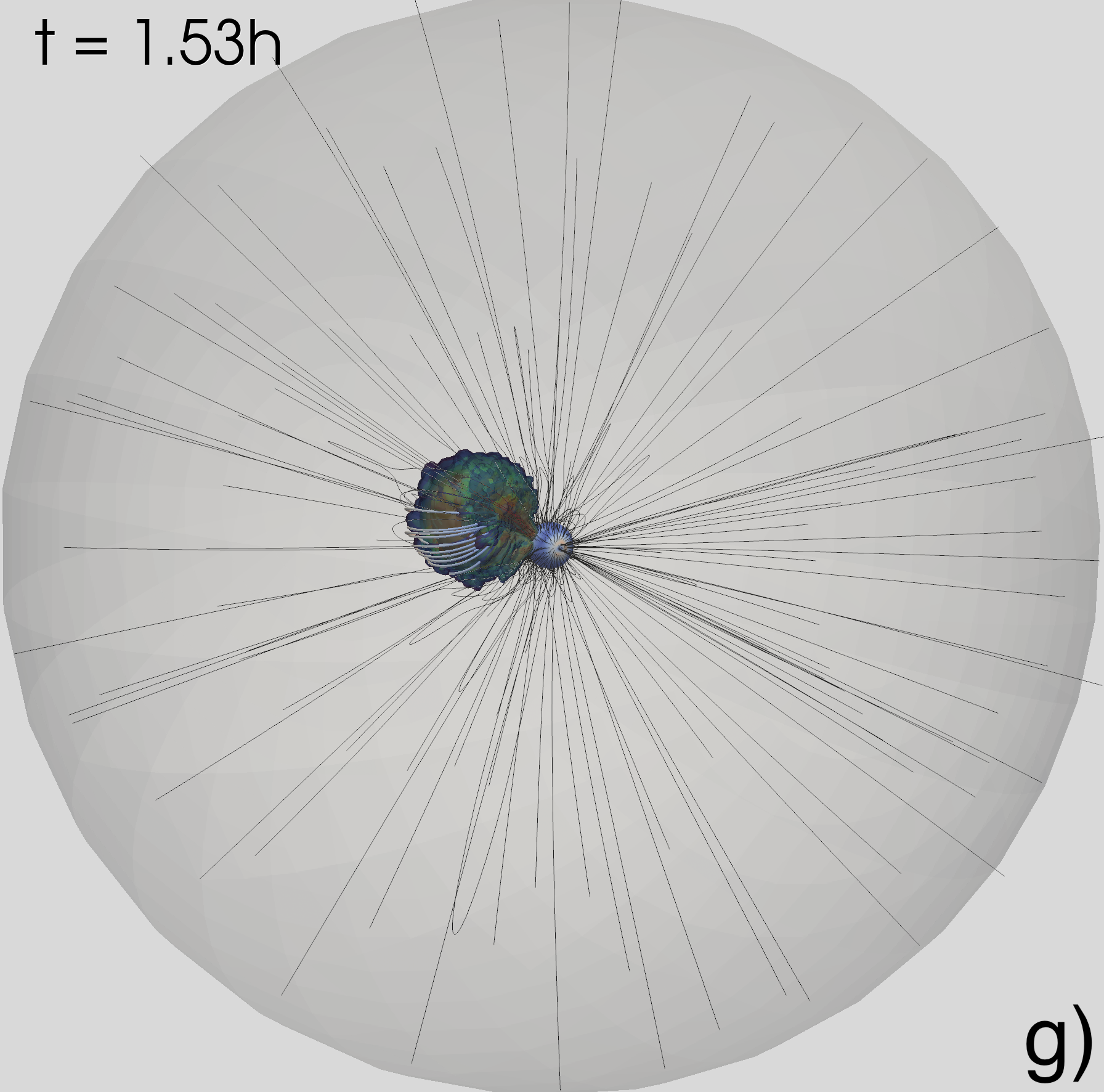}} \quad
    \subfigure{\includegraphics[width=0.30\textwidth]{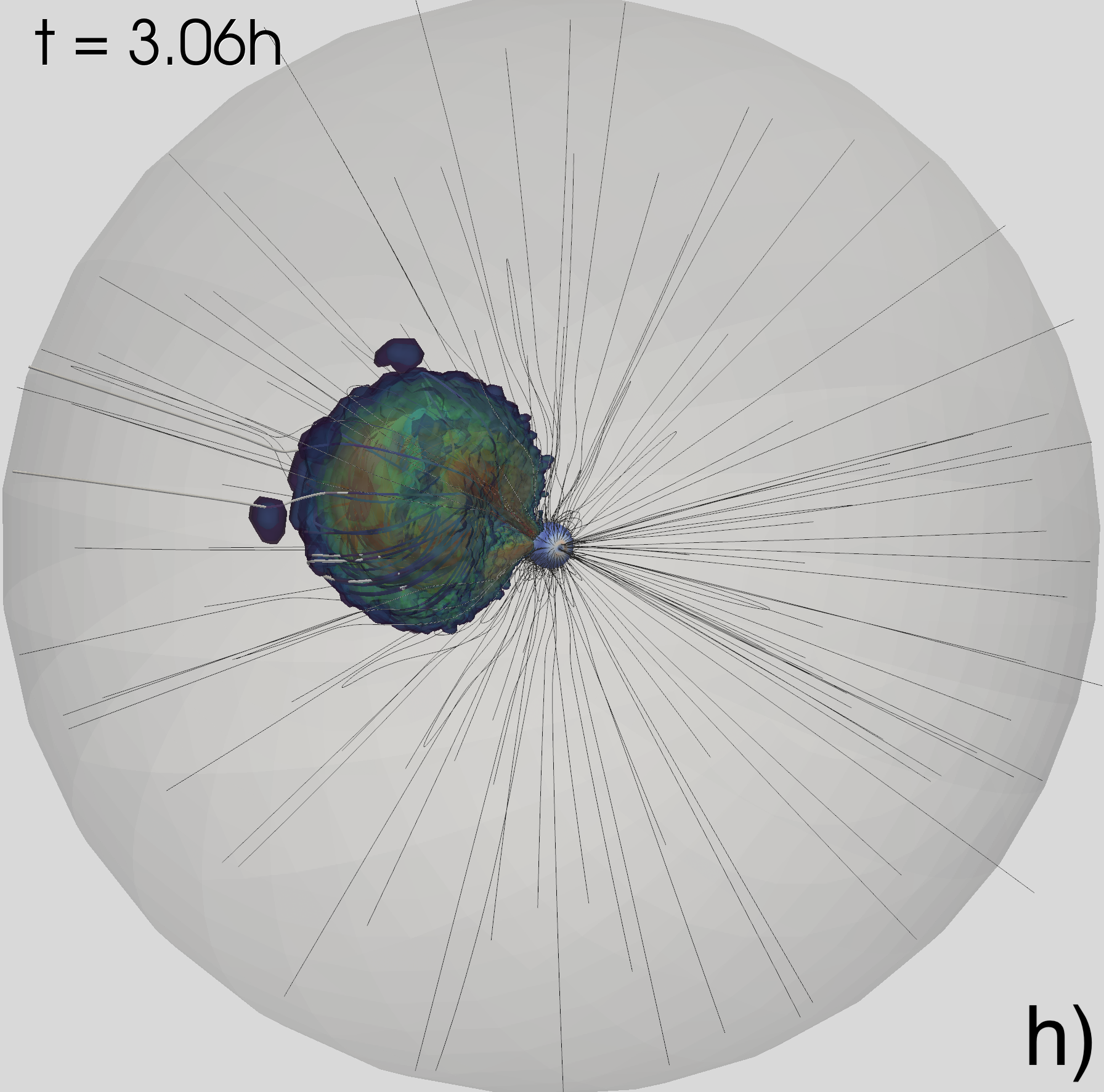}} \quad
    \subfigure{\includegraphics[width=0.30\textwidth]{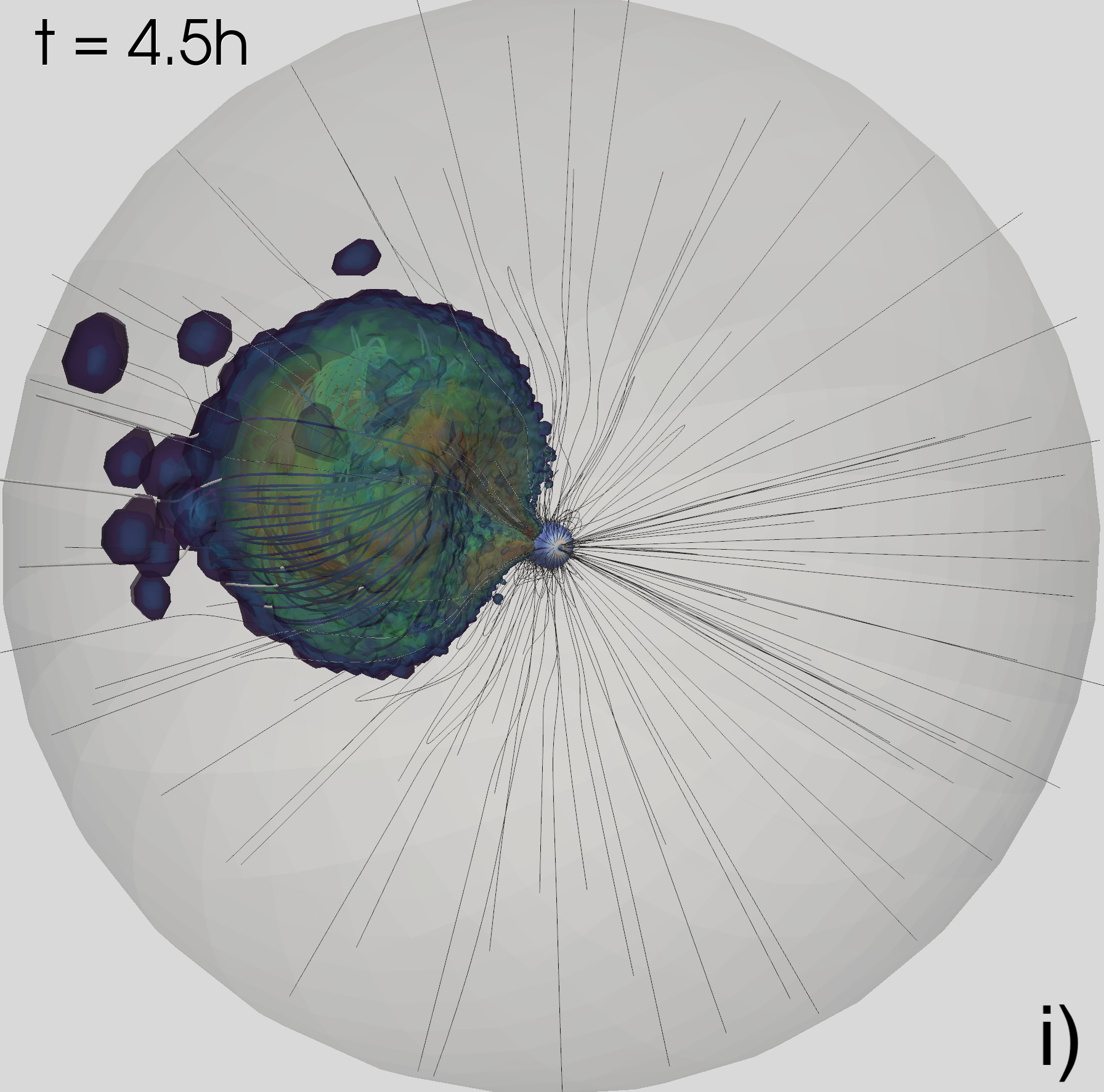}}
    \caption{Time-lapse of particle propagation in the flux rope with CFD using a perpendicular MFP depending on the Larmor radius of the test particle. The top panels (a through c) show the results for $\alpha = 10$, the middle panels (d through f) for $\alpha = 5$, and the bottom panels (g through i) for $\alpha = 1$. The same color bar as in Fig.~\ref{fig:no_cfd} is used for the intensities.}
    \label{fig:larmor}
\end{figure*}

Finally, we demonstrate simulation results, where CFD is implemented into the FTE by using $\lambda_\perp = \lambda_\perp(\alpha, \lambda_\parallel, r_\mathrm{L})$ that depends in particular on the Larmor radius $r_\mathrm{L}$ of the particle, since $\lambda_\parallel$ is kept constant. Different $\alpha$-values are prescribed (see Sec.~\ref{app:paradise} and Eq.~\ref{eq:lambda_perp_larmor}). The columns in Fig.~\ref{fig:larmor} are arranged similarly to those in Fig.~\ref{fig:const_mfp_perp}, but here, the top, middle, and bottom row display results using $\alpha = 10$, $\alpha = 5$, and $\alpha = 1$, respectively. Since $\lambda_\perp$ is inversely proportional to the magnetic field strength, $\lambda_\perp$ becomes in our simulations very large (about one order of magnitude larger than $\lambda_\parallel$) in regions containing a current sheet, where the magnetic field strength is very small. For this reason, we limited $\lambda_\perp$ to a maximum value of $\lambda_\parallel$. That is, in the center of the current sheets, we prescribe an isotropic spatial diffusion coefficient.

In the top row with $\alpha = 10$, we observe the most significant amount of CFD, even stronger than in the case of a constant $\lambda_\perp = 2.150 \times 10^{-2}\,R_\odot$ in Fig.~\ref{fig:const_mfp_perp}, top row. Already in panel a) of Fig.~\ref{fig:larmor}, particles stream away from the CME in all directions, while in panels b) and c), the half-plane containing the flux rope is nearly filled with the test particles. Furthermore, panels b) and c) showcase that particles move increasingly into the opposite half-circle. 

Decreasing $\alpha$ to a value of 5, as displayed in the panels of the middle row, results in a significant reduction of CFD compared to the prior case but still shows widespread test particles. In panel d), particles already tend to move away from the TDFR in all directions, while in panels e) and f), the test particles further fill out the half-circle into which the CME is propagating.

Reducing $\alpha$ further to 1, as illustrated in the bottom row, the least amount of CFD among the three presented cases is noticed. In panels g) and h), particles are still mainly confined to the CME along the interior and exterior magnetic field lines. Panel i) is comparable to the case of constant $\lambda_\perp = 2.150 \times 10^{-3}\,R_\odot$ in Fig.~\ref{fig:const_mfp_perp}, showing that particles still do not spread as widely as in the previous cases but receive access to the open field lines at the flux rope nose. 

\begin{figure*}[t]
    \centering
    \subfigure{\includegraphics[width=0.30\textwidth]{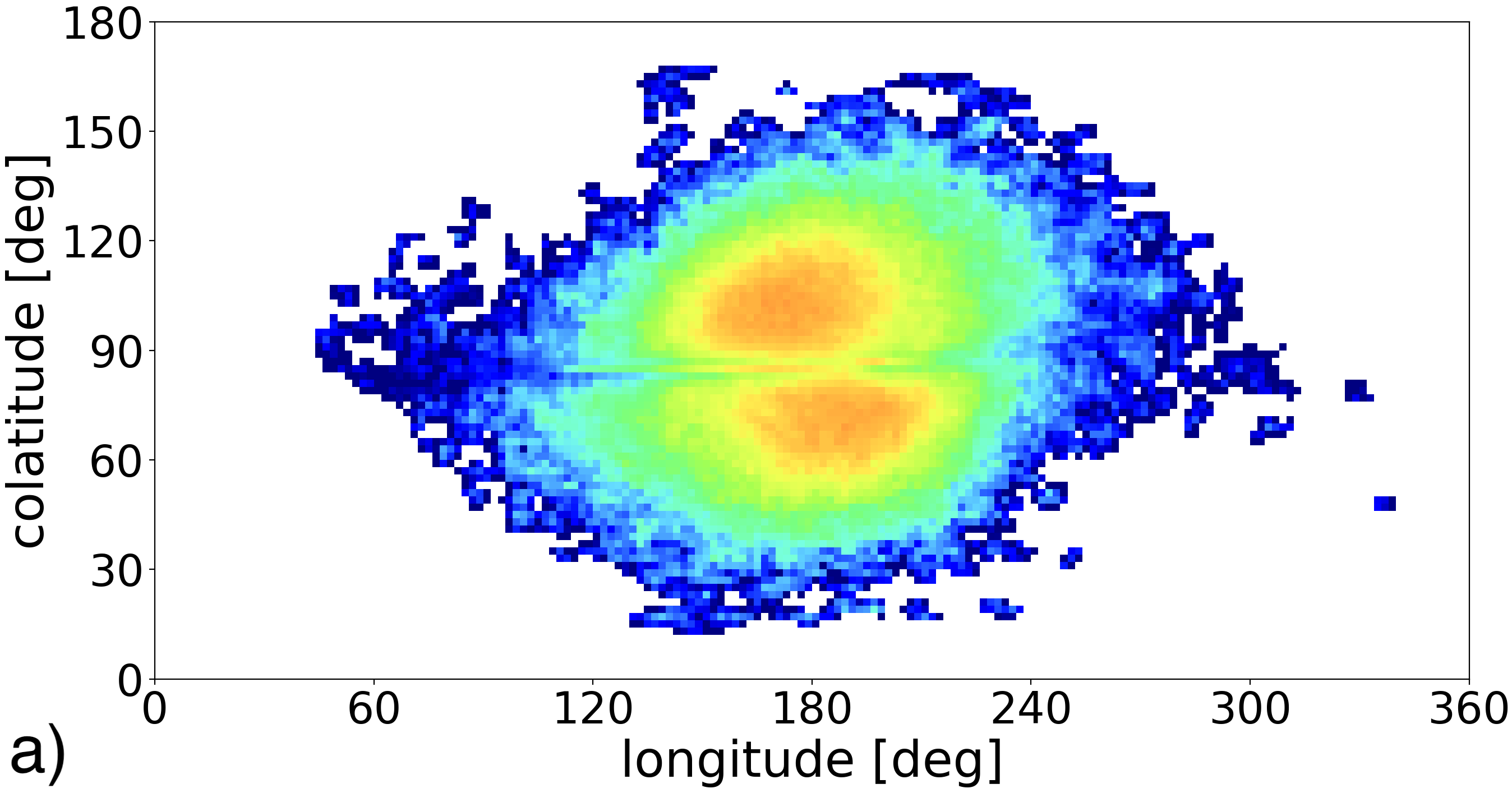}} \quad
    \subfigure{\includegraphics[width=0.30\textwidth]{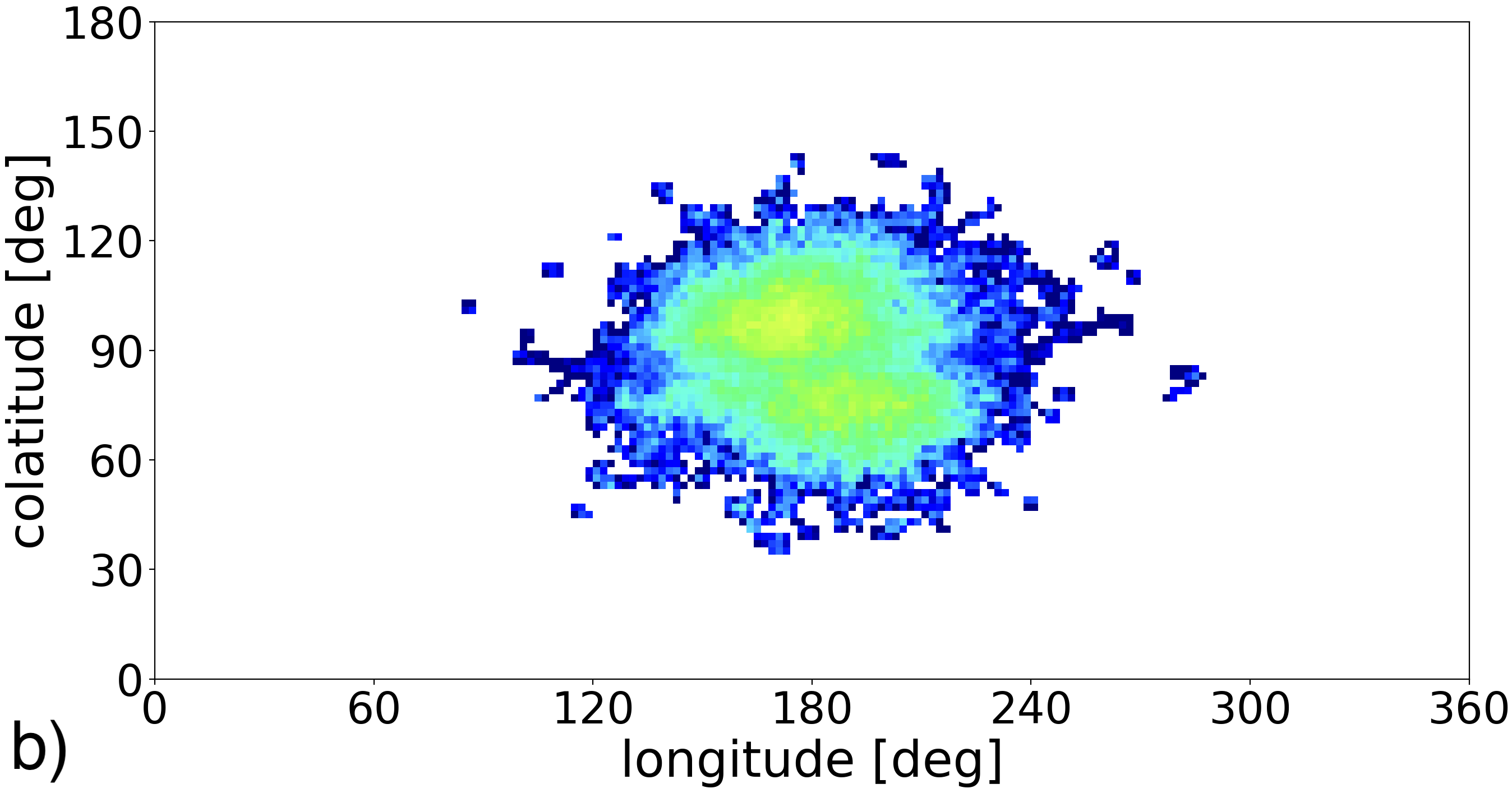}} \quad
    \subfigure{\includegraphics[width=0.30\textwidth]{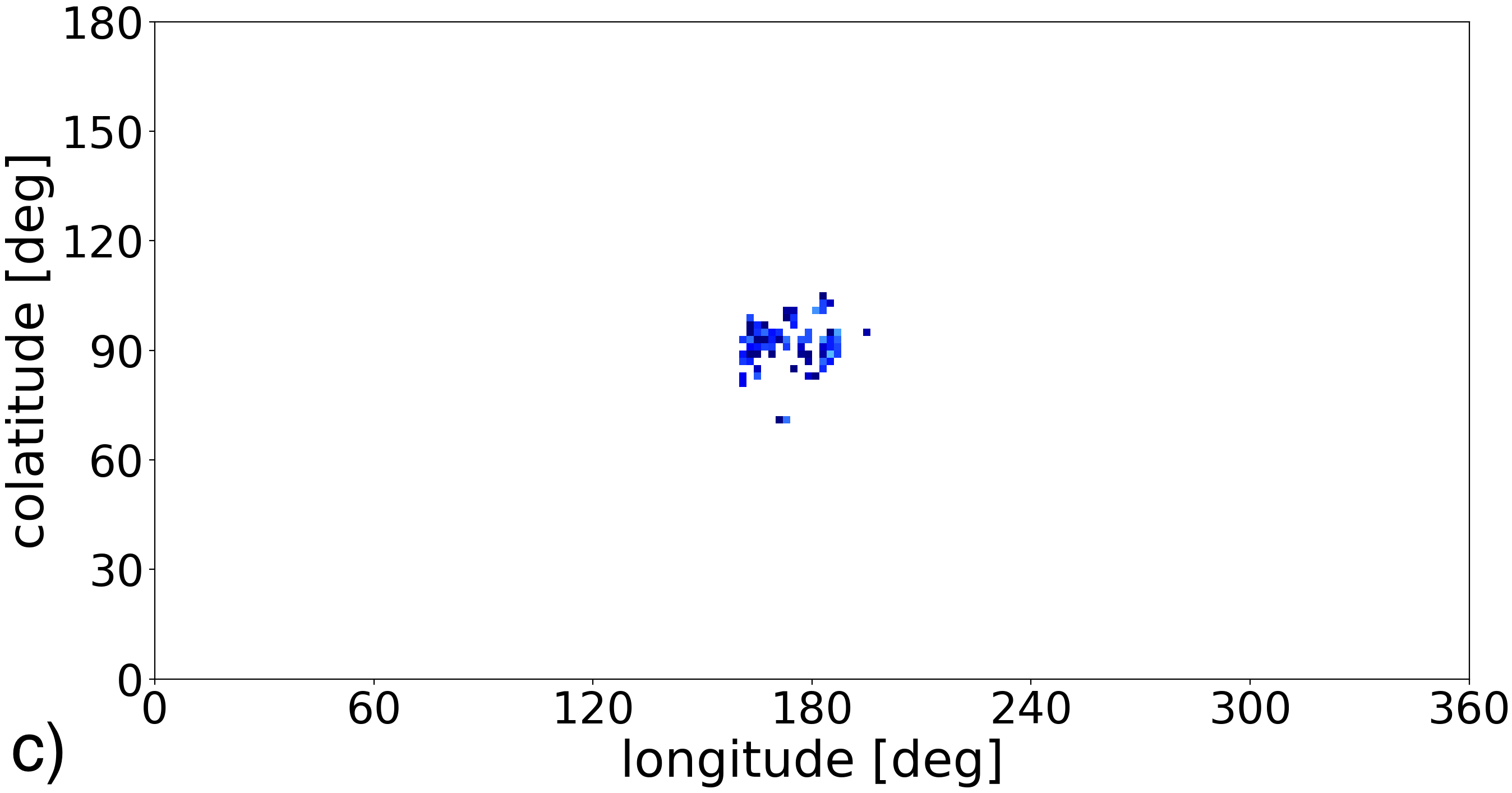}} \\
    \subfigure{\includegraphics[width=0.30\textwidth]{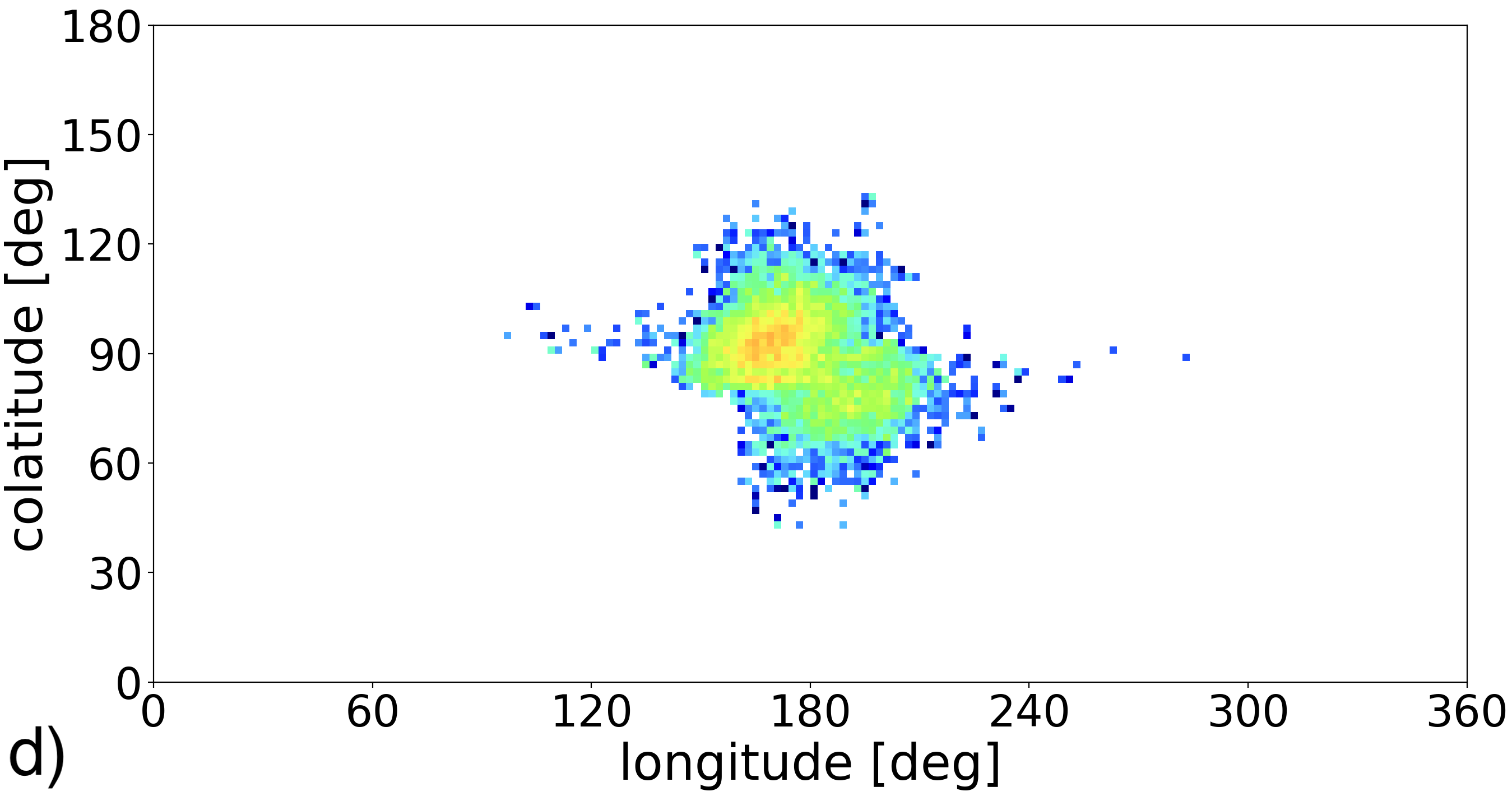}} \quad
    \subfigure{\includegraphics[width=0.30\textwidth]{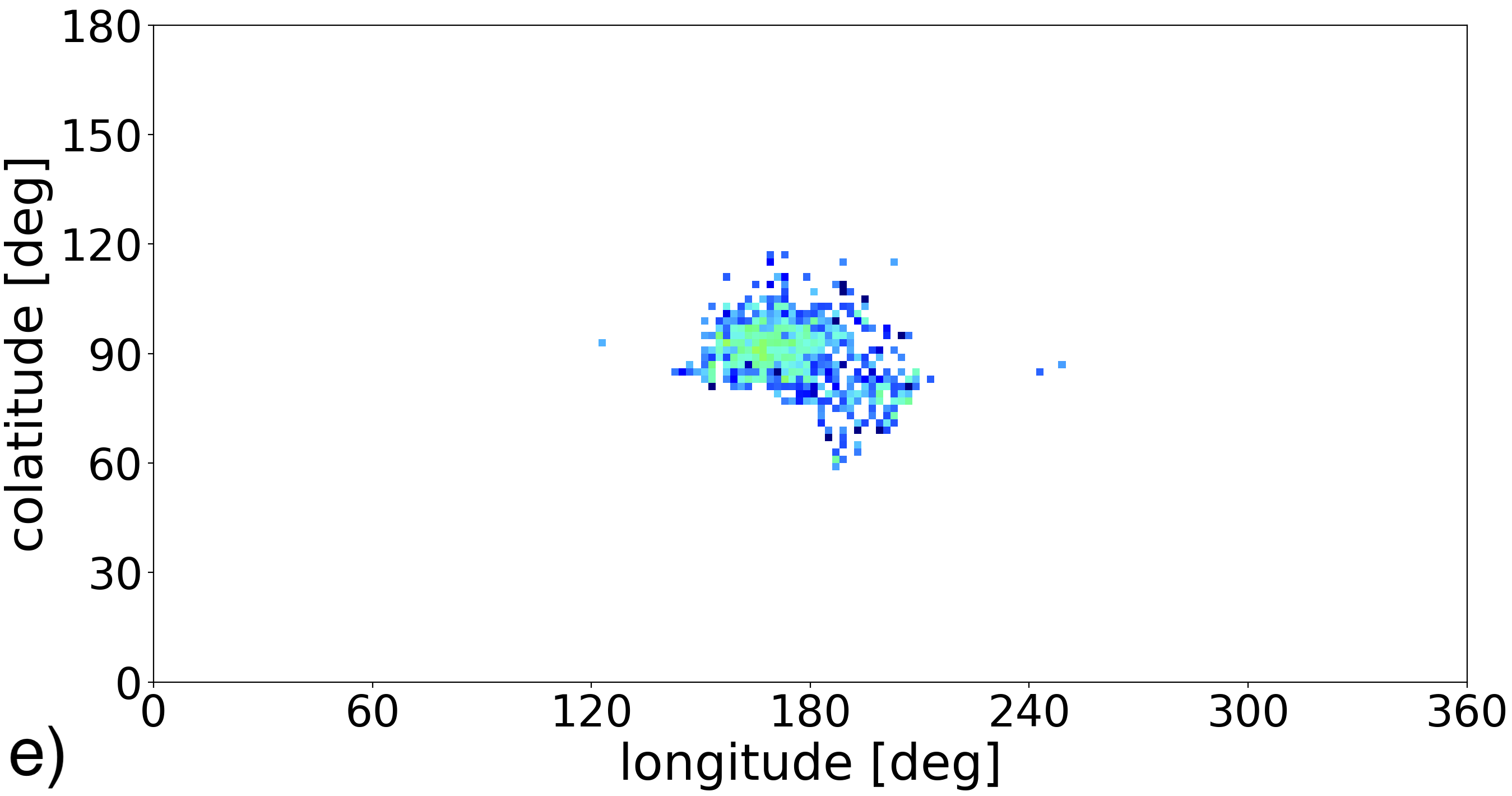}} \quad
    \subfigure{\includegraphics[width=0.30\textwidth]{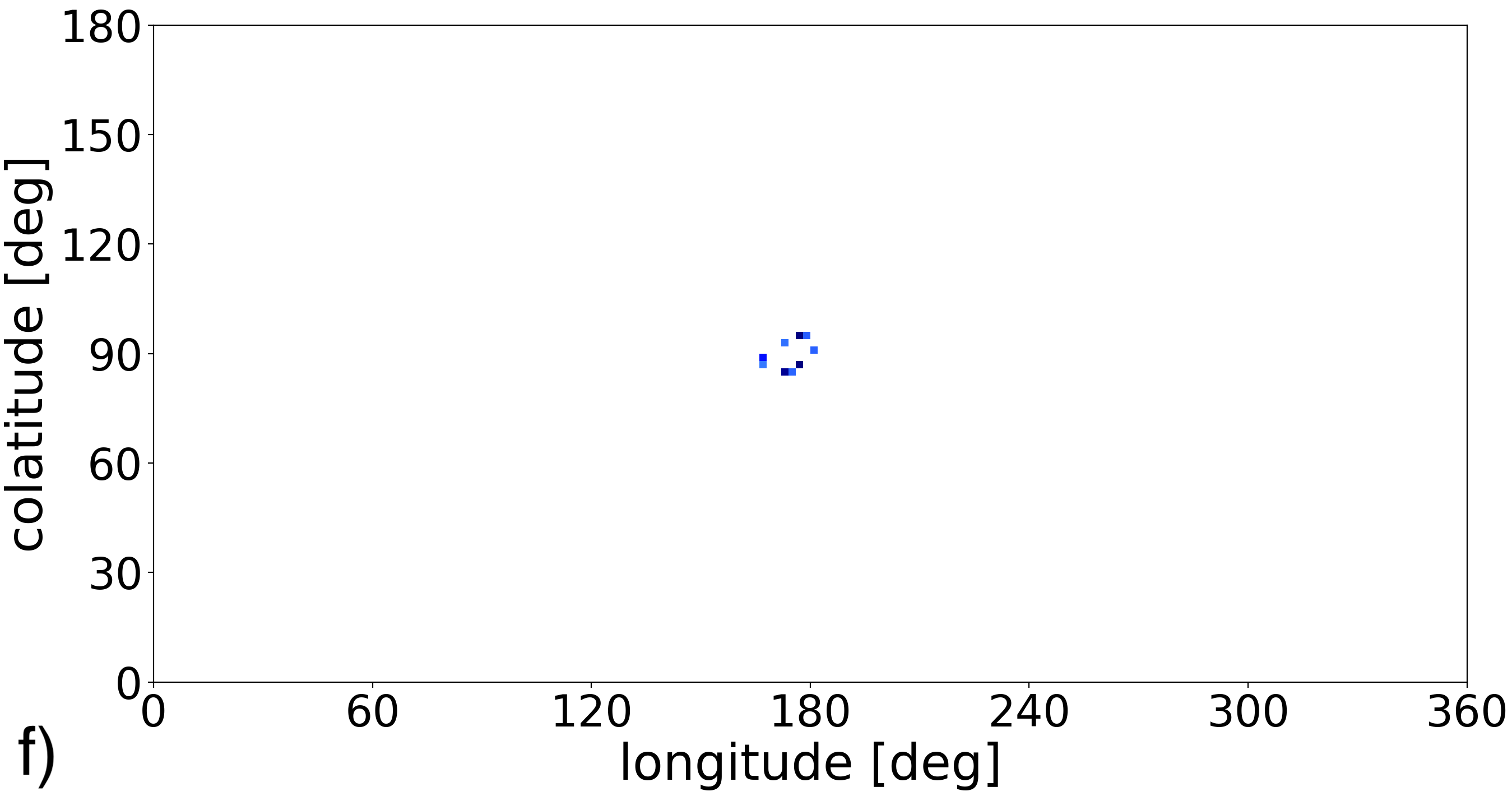}}
    \begin{minipage}[b]{0.8\textwidth}
        \includegraphics[width=\textwidth]{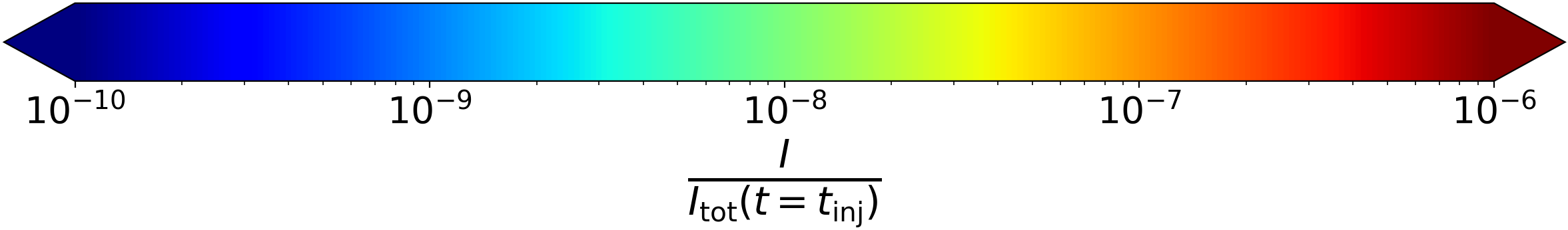} 
    \end{minipage}
    \caption{Intensity contour plots illustrating particle escape from the flux rope. The 2D plots show the particle intensities at the outer boundary of the domain. The top row contains the results for the Larmor radius-dependent $\lambda_\perp$, where $\alpha = 10$ in panel a), $\alpha = 5$ in panel b), and $\alpha = 1$ in panel c). The bottom row contains the results for the model with $\lambda_\perp = \mathrm{const}$, where $\lambda_\perp = 2.150 \times 10^{-2}\,R_\odot$ in panel d), $\lambda_\perp = 1.075 \times 10^{-2}\,R_\odot$\,au in panel e), and $2.150 \times 10^{-3}\,R_\odot$ in panel f).}
    \label{fig:I_integrated}
\end{figure*}

To quantify the number of particles that escaped the CME in the simulations with CFD, we integrated the intensities at the outer boundary of the simulation domain from $t = 0$\,h to $t = 6.5$\,h (a time short before the nose of the CME reaches $21.5\,R_\odot$). The 2D intensity contour plots in the $\theta-\phi$ plane are showcased in Fig.~\ref{fig:I_integrated}. The top row contains the plots for the Larmor radius-dependent model ($\alpha = 10;5;1$ in panels a through c, respectively), while the bottom row displays the plots for a constant $\lambda_\perp$ ($\lambda_\perp = 2.150 \times 10^{-2};1.075 \times 10^{-2};2.150 \times 10^{-3}\,R_\odot$ in panels d through f, respectively). The largest spread of particles and the most significant number of escaped particles occurred in the Larmor radius-dependent model with $\alpha = 10$, where most of the particles escaped in the direction parallel to the legs of the TDFR. Even the case with $\alpha = 5$ results in a broader spread of particles compared to all cases with constant $\lambda_\perp$. In general, Fig.~\ref{fig:I_integrated} illustrates that in all CFD simulations, the most substantial effect of CFD is noticeable in the propagation direction of the CME (evident in the most significant intensities being around $90^\circ$ colatitude and $180^\circ$ longitude.) As suggested from the plots in Figs.~\ref{fig:const_mfp_perp} and \ref{fig:larmor}, the cases with the smallest parameter values in both models show the least amount of escaped particles and smallest spread of particles.

\section{Summary and Outlook} \label{sec:summary}

This paper introduced the novel COCONUT\allowbreak$+$\allowbreak PARADISE model aimed at simulating the acceleration and transport of energetic particles in the solar corona. Using the global coronal 3D MHD model COCONUT, we generated coronal background configurations containing a modified TDFR CME. Subsequently, we employed the particle transport code PARADISE to evolve energetic particles as test particles through these backgrounds to investigate the general propagation of particles within the TDFR and, in particular, the effects of CFD on particle transport in the corona by using two different approaches for the CFD coefficient in the FTE. In all simulations, we injected monoenergetic 100\,keV protons in one of the legs of the TDFR close to its footprint. The simulation without any CFD mechanism showed that particles initially spread along the interior field lines of the TDFR, later gaining access to the TDFR's exterior field lines. However, the particles remained confined to the CME and did not gain access to the field lines ahead of the CME that had opened due to MR.

The first approach for the CFD coefficient assumed a constant perpendicular MFP $\lambda_\perp$. Even a relatively small value typically used in heliospheric simulations that employ this type of CFD model, such as $\lambda_\parallel = 21.5\,R_\odot$ and $\lambda_\perp / \lambda_\parallel  = 10^{-3}$, led to a large spread of particles along the longitudinal range. In the second approach to model CFD, we used a $\lambda_\perp$ dependent on the particle's Larmor radius and found qualitatively similar results as for the first approach. In both models, by reducing the constant $\lambda_\perp$ or the $\alpha$-parameter, respectively, within one order, we observed considerably less CFD. However, in all three presented cases, particles gained access to the reconnected open magnetic field lines at the nose of the TDFR and, as a result, escaped the flux rope predominantly in the propagation direction of the CME. 

As discussed in Sec.~\ref{sec:introduction}, observations by PSP, along with phenomena such as Type IV radio bursts and Forbush decreases, indicate that CMEs trap particles within their flux ropes and prevent external particles from entering the flux rope. The rapid increases in SEP intensity recorded by PSP in the solar corona during the massive September 2022 SEP event inside a flux rope suggest limited perpendicular diffusion in and around the CME. Consequently, the simulations without CFD or with weak CFD would be more representative of reality, demonstrating that particles can remain trapped in a flux rope for an extended period. However, our simulations reveal that the effect of CFD is most pronounced at the nose of the CME. Since PSP passed through the leg and backside of the CME during the September 2022 SEP event, it may thus have missed any strong CFD signatures. 

Several projects using the COCONUT$+$PARADISE model are planned. COCONUT is being integrated with the EUHFORIA heliospheric MHD model \citep{Pomoell-Poedts-2018} to simulate CME propagation from the lower corona to 1\,au and beyond. This will help extend our current study on particle confinement within magnetic flux ropes and investigate widespread SEP events. Such studies could extend this paper's solar minimum-based coronal simulations by modeling solar maximum conditions, where increased interaction between energetic particles, CMEs, and the ambient solar wind is expected.
Furthermore, in the present work, we neglected the minor acceleration for particles that escape the CME and cross the shock driven by the CME, and we 
will conduct a more detailed analysis of particle acceleration with the fully coupled EUHFORIA+COCONUT model.
We also aim to explore particle guiding center drifts, which are essential for understanding perpendicular transport \citep[e.g.,][]{vandenBerg-etal-2021}. Additionally, the combined COCONUT$+$PARADISE model can be used to investigate the back-propagation of energetic particles during long-duration gamma-ray flares \citep[e.g.,][]{Hutchinson-etal-2022}. Future work will benefit from observations by various spacecraft, especially PSP, which will provide valuable in-situ data to validate and refine our simulations, thereby enhancing our understanding of particle dynamics in the solar environment. Finally, we plan to compare particle transport in complex flux ropes (such as the TDFR) with that in static, twisted and untwisted coronal loops, to explore the differences in particle confinement and CFD between these magnetic structures, offering deeper insights into the effects of magnetic twist and evolving magnetic fields.

\section*{Acknowledgments}

E.H.\ is grateful to the Space Weather Awareness Training Network (SWATNet), funded by the European Union's Horizon 2020 research and innovation program under the Marie Skłodowska-Curie grant agreement No.\ 955620. Furthermore, E.H.\ acknowledges the travel grant V477923N from the Fonds voor Wetenschappelijk Onderzoek – Vlaanderen (FWO).
N.W.\ acknowledges funding from the Research Foundation -- Flanders (FWO -- Vlaanderen, fellowship no.\ 1184319N).
S.P.\ acknowledges support from the projects C16/24/010  (C1 project Internal Funds KU Leuven), G0B5823N and G002523N (WEAVE)   (FWO-Vlaanderen), 4000145223 (SIDC Data Exploitation (SIDEX2), ESA Prodex), and Belspo project B2/191/P1/SWiM. This project has received funding from the European Research Council Executive Agency (ERCEA) under the ERC-AdG agreement No 101141362 (Open SESAME).
Computational resources and services used in this work were provided by the VSC (Flemish Supercomputer Center), funded by the Research Foundation - Flanders (FWO) and the Flemish Government – department EWI.

\appendix

\section{COCONUT}\label{app:coconut}

The coronal MHD model COCONUT, based on the COOLFLuiD platform (see Sec.~\ref{sec:introduction} for references), computes solar corona configurations by solving the ideal 3D MHD equations with gravity using a time-implicit backward Euler scheme. This system of MHD equations reads in conservative form as:
\begin{align}
    \frac{\partial}{\partial t}
    \begin{pmatrix}
        \rho \\
        \rho\,\vect{v} \\
        E \\
        \vect{B} \\
        \psi
    \end{pmatrix}
    + \nabla \cdot 
    \begin{pmatrix}
        \rho\,\vect{v} \\
        \rho\,\vect{v}\vect{v} + \mathcal{I}\left(
        P + \frac{1}{2} \vert \vect{B}\vert^2 \right) 
        - \vect{B}\vect{B}\\
        \left(E + P + \frac{1}{2} 
        \vert \vect{B}\vert^2 \right) \vect{v} 
        - \vect{B}\left(\vect{v} \cdot \vect{B}\right)\\
        \vect{v}\vect{B} - \vect{B}\vect{v} + 
        \mathcal{I}\psi\\
        V_\mathrm{ref}^2\,\vect{B}
    \end{pmatrix}
    = 
    \begin{pmatrix}
        0 \\
        \rho\,\vect{g} \\
        \textbf{0} \\
        \rho\, \vect{g} \cdot \vect{v} \\
        0
    \end{pmatrix}\,. \label{eq:MHD_equations}
\end{align}
In Eq.~\eqref{eq:MHD_equations}, $\rho$ is the mass density of the background solar wind, $\vect{v}$ the plasma bulk velocity, $\vect{B}$ the background magnetic field vector, $P$ the thermal gas pressure (in units of energy), and $V_\mathrm{ref}$ a reference speed. The total energy $E$ is defined as $E = \rho\,\vect{v}^2 + \rho\,\mathcal{E} + \vect{B}^2/(8\,\pi)$ with internal energy $\mathcal{E}$, while the gravitational acceleration is given by $\vect{g}(r) = -(G\,M_\odot/r^2)\,\hat{\vect{e}}_r$, where $G$ denotes the gravitational constant, $M_\odot$ the solar mass, $r$ the radial distance from the Sun, and $\hat{\vect{e}}_r$ the unit vector in the radial direction. The variable $\mathcal{I}$ denotes the identity dyadic which, in terms of the canonical unit vectors, is expressed as $\mathcal{I} = \hat{\vect{e}}_x \otimes \hat{\vect{e}}_x + \hat{\vect{e}}_y \otimes \hat{\vect{e}}_y + \hat{\vect{e}}_z \otimes \hat{\vect{e}}_z$. The set of ideal MHD equations in the first four rows of Eq.~\eqref{eq:MHD_equations}, that is, the continuity, momentum, energy, and Faraday's induction equation, respectively, are accompanied by a complementary equation containing the Lagrange multiplier $\psi$. This equation serves to numerically ensure the solenoidal constraint $\nabla \cdot \vect{B} = 0$ \citep[for details, see][]{Perri-etal-2022}. To close the system of equations in Eq.~\eqref{eq:MHD_equations}, we consider a polytropic process and use the ideal equation of state $\rho\,\mathcal{E} = P/(\gamma -1)$ with a reduced adiabatic index of 1.05.

A major difference between the MHD models previously used with PARADISE and the new model COCONUT is that the former models employ a structured grid. At the same time, COCONUT solves the MHD equations on an unstructured grid consisting of prisms with equilateral triangular faces (see \citealt{Brchnelova-etal-2022} for details). Consequently, we implemented alternative methods in PARADISE to trace the particles and calculate the gradients of the velocity and magnetic field components on an unstructured grid. Hitherto, these procedures take significantly longer than those in the original PARADISE architecture. While we continue to improve the performance of PARADISE using the original COCONUT output, we have also interpolated the original unstructured COCONUT grid to a structured spherical grid, allowing us to utilize the faster original PARADISE architecture. The original unstructured COCONUT grid is employed in the paper, and a comparison between the unstructured and interpolated structured grid is presented in Appendix~\ref{app:interpolation}. While COCONUT itself has been benchmarked against other models and observations \citep{Perri-etal-2022, Perri-etal-2023, Kuzma-etal-2023}, the interpolation of COCONUT to a EUHFORIA-like grid serves as a test for the correct implementation of the unstructured grid in PARADISE.

\section{PARADISE}\label{app:paradise}

The PARADISE model simulates the acceleration and transport of energetic particles embedded in a background plasma provided by various MHD models (see Sec.~\ref{sec:introduction}). The spatio-temporal particle intensity distributions are obtained by solving the time-dependent focused transport equation \citep[FTE; e.g.,][]{Roelof-1969, Skilling-1971, Skilling-1975,Ruffolo-1995, Isenberg-1997, le-Roux-Webb-2009}. Between two consecutive MHD snapshots, a linear interpolation in time is performed. Unlike particle transport models such as EPREM and M-FLAMPA (see Sec.~\ref{sec:introduction}) that use a finite difference method for solving the FTE, PARADISE solves a set of stochastic differential equations (SDEs) that is equivalent to the FTE, by applying It$\hat{\mathrm{o}}$ calculus (see, e.g., \cite{Strauss-Effenberger-2017} for an overview). Using SDEs allows for simple parallelization of the computation by dividing the total number of pseudo-particles over multiple cores.

The FTE utilized in PARADISE can be compactly written as
\begin{align}
    \frac{\partial f}{\partial t} 
    + \frac{\diff\vect{x}}{\diff t} \cdot \nabla f 
    + \frac{\diff \mu}{\diff t} 
    \frac{\partial f}{\partial \mu} 
    + \frac{\diff p}{\diff t} 
    \frac{\partial f}{\partial p}
    = \frac{\partial}{\partial \mu} 
    \left(D_{\mu\mu} \frac{\partial f}{\partial \mu} \right)
    + \nabla \cdot \left( \vect{\kappa}_\perp
    \cdot \nabla f \right) \label{eq:FTE}\,.
\end{align}
Here, $f = f(\vect{x}; p, \mu, t)$ is a five-dimensional gyrotropic particle distribution function as a function of spatial coordinates $\vect{x}$, momentum magnitude $p$, pitch angle cosine $\mu \equiv \cos(\beta)$, and time $t$. The pitch angle $\beta \equiv \arctan(v_\perp/v_\parallel)$ is defined as the angle between the velocity components perpendicular ($v_\perp$) and parallel ($v_\parallel$) to the ambient mean magnetic field. Turbulence in the solar wind leading to diffusion processes in phase-space is modeled by including the pitch angle diffusion coefficient $D_{\mu\mu}$ and the spatial cross-field diffusion tensor $\vect{\kappa}_\perp$. The total derivatives of $\vect{x}$, $\mu$, and $p$ in Eq.~\eqref{eq:FTE} read in detail
\begin{align}
    \frac{\diff \vect{x}}{\diff t} 
    &= \vect{V}_\mathrm{sw} + \vect{V}_\mathrm{d} + 
    \mu\,v\,\vect{b} \label{eq:dxdt}\,, \\
    \frac{\diff \mu}{\diff t} 
    &= \frac{1 - \mu^2}{2} \left(v\,\nabla \cdot \vect{b} 
    + \mu\,\nabla \cdot \vect{V}_\mathrm{sw}
    - 3\,\mu\,\vect{b}\vect{b}:\nabla \vect{V}_\mathrm{sw}
    - \frac{2}{v}\vect{b} \cdot 
    \frac{\diff \vect{V}_\mathrm{sw}}{\diff t}\right)\,,
    \label{eq:dmudt} \\
    \frac{\diff p}{\diff t} &= 
    \left[\frac{1 - 3\,\mu^2}{2} 
    \left(\vect{b}\vect{b}:\nabla \vect{V}_\mathrm{sw}\right)
    - \frac{1-\mu^2}{2} \nabla \cdot \vect{V}_\mathrm{sw}
    - \frac{\mu}{v} \vect{b} \cdot 
    \frac{\diff \vect{V}_\mathrm{sw}}{\diff t}\right]p\,.
    \label{eq:dpdt}
\end{align}
In Eq.~\eqref{eq:dxdt}, $\vect{V}_\mathrm{sw}$ is the plasma bulk velocity of the ambient solar wind, $\vect{V}_\mathrm{d}$ comprises various drift velocities of a particle's guiding centre, $v = p/(\gamma\,m)$ describes the particle's speed using the Lorentz factor $\gamma$ and its rest mass $m$, and $\vect{b}$ denotes the unit vector parallel to the ambient mean magnetic field. The colon in Eqs.~\eqref{eq:dmudt} and \eqref{eq:dpdt} stands for the Frobenius inner product $\vect{b}\vect{b}:\nabla \vect{V}_\mathrm{sw} = b_{ij}\, \partial v_i / \partial x_j$ (using Einstein's summation convention).

The pitch angle diffusion coefficient in the FTE is derived from quasi-linear theory, and, following \citet{Agueda-Vainio-2013} and \citet{Wijsen-etal-2019a}, implemented in the form
\begin{align}
    D_{\mu\mu} = D_0 \left(\frac{R}{R_0}\right)^{2-d}\left(\frac{\vert \mu \vert}{1 
    + \vert \mu \vert} + \epsilon \right)
    \left(1 - \mu^2\right)\,.\label{eq:Dmumu}
\end{align}
where $d = 5/3$ is the Kolmogorov spectral index, $R$ is the particle rigidity, and the reference particle rigidity $R_0$ is for a proton with a reference energy of 1\,MeV. 
The additional parameter $\epsilon = 0.048$ is included to handle the resonance gap at $\mu = 0$ \citep{Klimas-Sandri-1971}. The scaling factor $D_0$ is related to the parallel mean free path $\lambda_\parallel^\mathrm{r}$ via \citep{Hasselmann-Wibberenz-1970}
\begin{align}
    \lambda_\parallel = \frac{3\,v}{8} 
    \int\limits_{-1}^{1} 
    \frac{(1-\mu^2)^2}{D_{\mu\mu}}\,\mathrm{d}\mu\,.\label{eq:lambda_r}
\end{align}

In general, to describe the CFD tensor in Eq.~\eqref{eq:FTE} we use, following \cite{Zhang-etal-2009} and \cite{Wang-etal-2012},
\begin{align}
    \vect{\kappa}_\perp = \kappa_\perp\,\left(\mathcal{I} - \vect{b}\vect{b}\right) \label{eq:CFD_tensor}
\end{align}
with $\vect{b}\vect{b}$ denoting the dyadic product of the magnetic field unit vector with itself, and $\kappa_\perp$ reading
\begin{align}
    \kappa_\perp = \lambda_\perp \frac{v}{3}\,.\label{eq:const_lambda_perp}
\end{align}
The first approach involves simply assuming a constant $\lambda_\perp$, and we use varying values for $\lambda_\perp$ to investigate the effect of CFD on the particle transport. For the second approach, $\lambda_\perp$ becomes a function of $\lambda_\parallel$, a scaling factor $\alpha$ that is a measure of the ratio of $\lambda_\perp$ to $\lambda_\parallel$, and, importantly, the particle's maximum Larmor radius
\begin{align}
    r_\mathrm{L} = \frac{m\,v}{\vert q \vert\,B}\,. \label{eq:larmor_radius}
\end{align}
with charge $q$. In this case, $\kappa_\perp$ reads
\begin{align}
    \kappa_\perp = \underbrace{\frac{3}{4}\,\pi\,\alpha \frac{r_\mathrm{L}}{r_\mathrm{L,0}} \lambda_\parallel}_{=\,\,\lambda_\perp}\,v\,. \label{eq:lambda_perp_larmor}
\end{align}
In Eq.~\eqref{eq:lambda_perp_larmor}, $r_\mathrm{L,0}$ is a reference Larmor radius determined by a reference magnetic field strength $B_0$ and a reference energy (here, $E = 1$\,MeV). Furthermore, $r_\mathrm{L,0}$ is averaged over $\mu$ and thus is actually the maximum Larmor radius (corresponding to the case of $\mu = 0$). 

\begin{figure*}[ht]
    \centering
    \begin{minipage}[b]{0.48\textwidth}
        \centering
        \includegraphics[width=\textwidth]{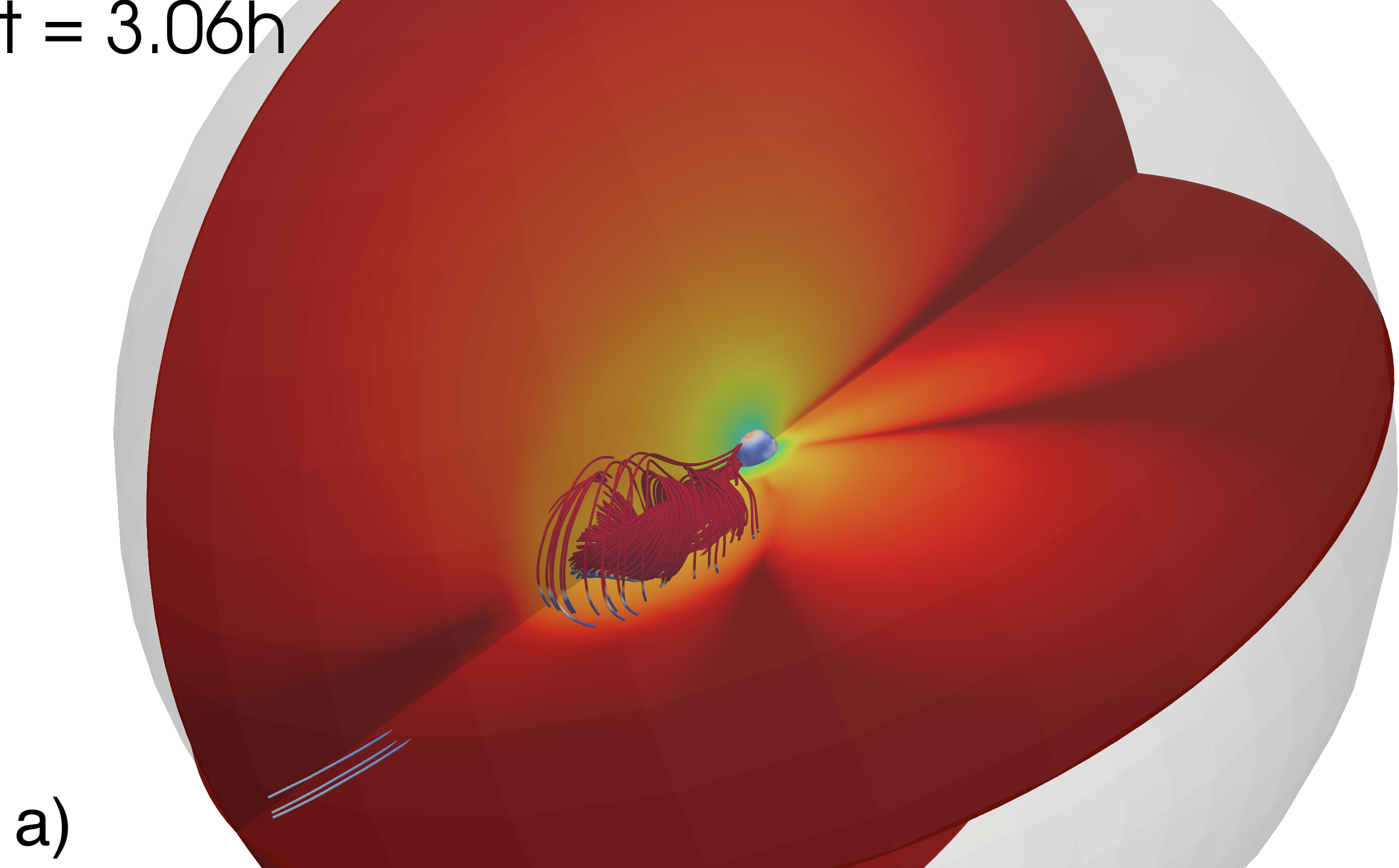}
        \vspace{1em} 
        \includegraphics[width=0.8\textwidth]{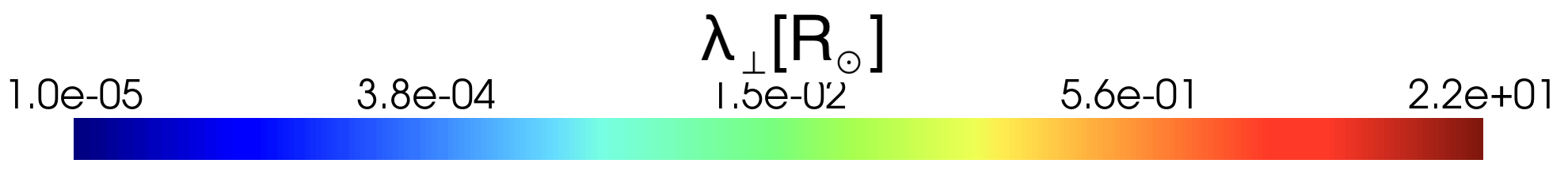}
    \end{minipage}%
    \quad 
    \begin{minipage}[b]{0.48\textwidth}
        \centering
        \includegraphics[width=\textwidth]{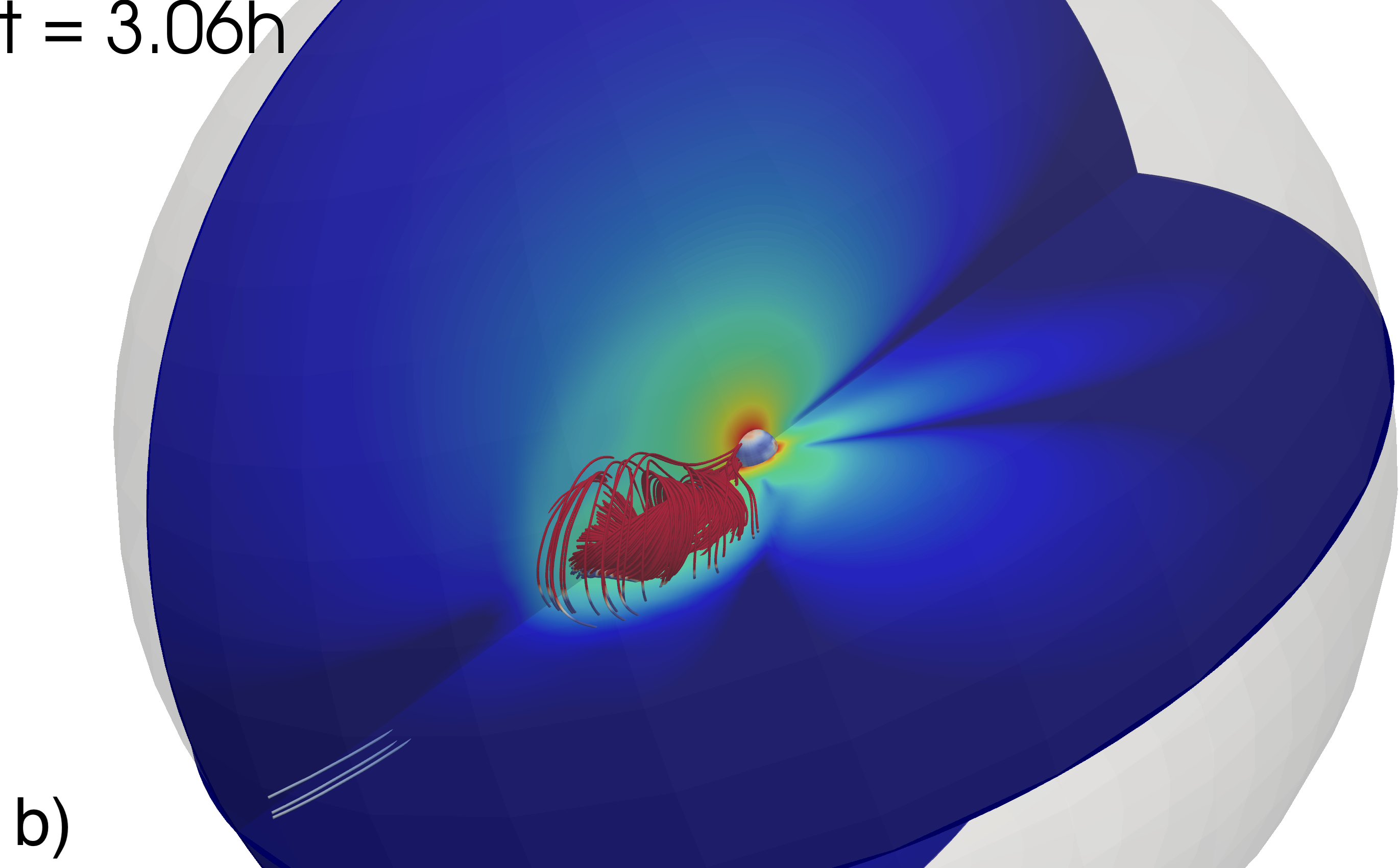}
        \vspace{1em} 
        \includegraphics[width=0.8\textwidth]{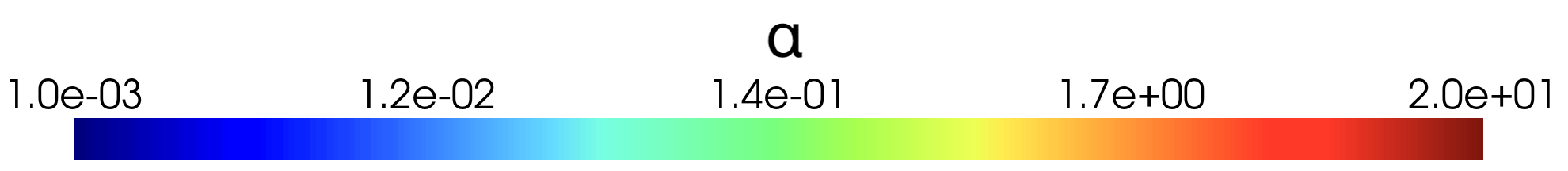}
    \end{minipage}
    
    \caption{Contour plots of the Larmor radius-dependent perpendicular MFP $\lambda_\perp$ (panel a) in the case of $\alpha = 5$ (corresponding to results shown in the middle row of Fig.~\ref{fig:larmor}) and of the $\alpha$-parameter (panel b) based on the case of a constant MFP of $1.075 \times 10^{-2}\,R_\odot$ (corresponding to the middle row of Fig.~\ref{fig:const_mfp_perp}). The contours are shown in the two planes perpendicular to the CME.}
    \label{fig:mfp}
\end{figure*}

To provide the reader with an understanding of the Larmor radius-dependent $\lambda_\perp$ and the $\alpha$ values in the case of constant $\lambda_\perp$, Fig.~\ref{fig:mfp} presents contour plots of these parameters in the two planes perpendicular to the CME. The left panel shows the Larmor radius-dependent $\lambda_\perp$ in the case of $\alpha = 5$ (corresponding to the middle row of Fig.~\ref{fig:larmor}, while the right panel showcases the $\alpha$ values based on the case of constant $\lambda = 1.075 \times 10^{-2}\,R_\odot$ (see middle row of Fig.~\ref{fig:const_mfp_perp}). In both cases, the Larmor radius (Eq.~\ref{eq:larmor_radius}) was calculated based on the speed of a proton of 100\,keV energy. The left panel shows distinct regions where $\lambda_\perp$ has comparatively large values, which would exceed those of $\lambda_\parallel$ by an order of magnitude if the limit were not set to $\lambda_\parallel$. These large $\lambda_\parallel$ values result from very small local magnetic field strengths due to current sheets. In the simulations discussed in Sec.~\ref{fig:larmor}, the level of CFD was largely influenced by the $\lambda_\perp$ values in and near the CME, rather than by the large $\lambda_\perp$ values along the CME propagation direction.

\section{Interpolating Unstructured COCONUT to a Structured Grid}\label{app:interpolation}

\begin{figure*}[t]
    \centering
    \subfigure{\includegraphics[width=0.40\textwidth]{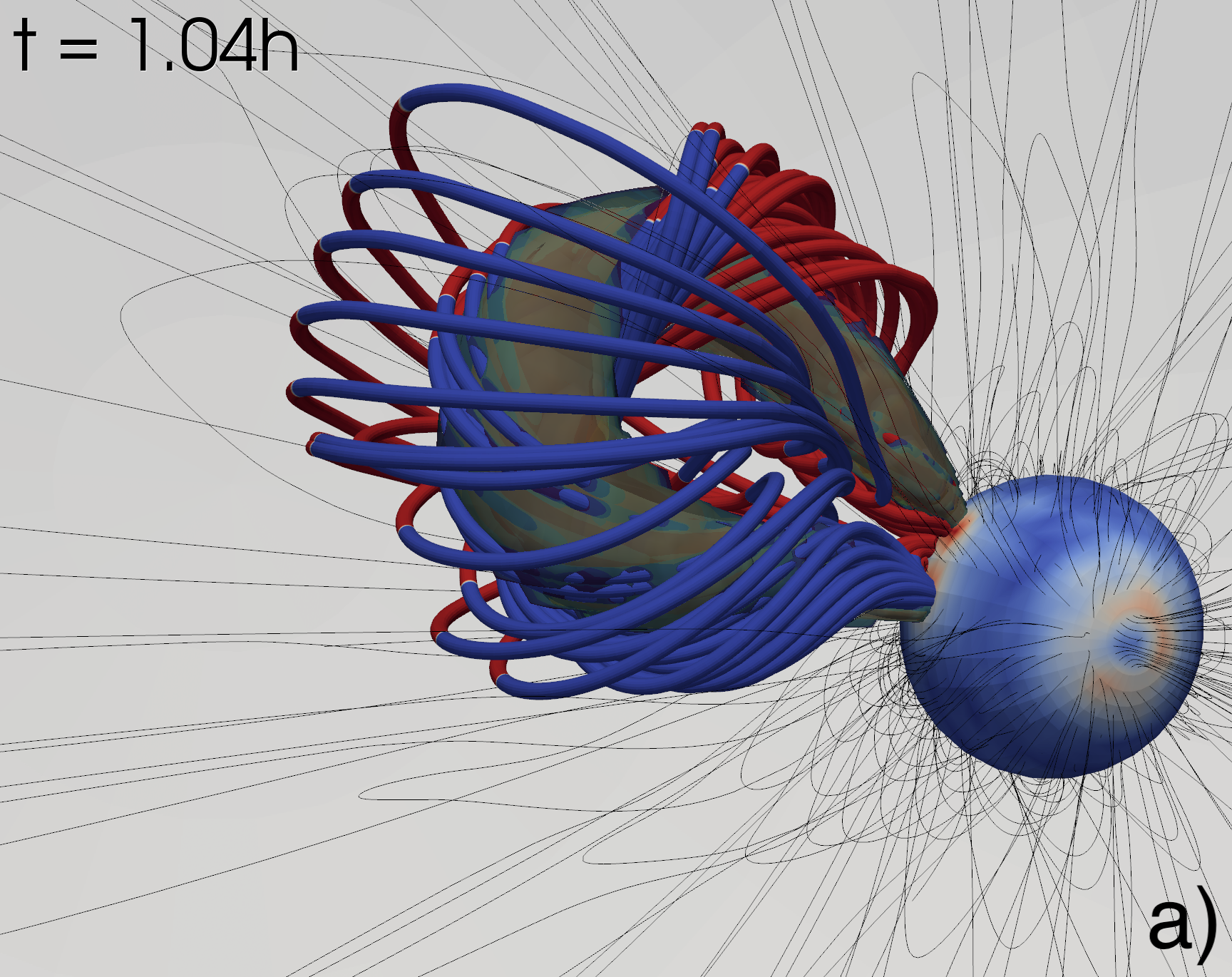}} \quad
    \subfigure{\includegraphics[width=0.40\textwidth]{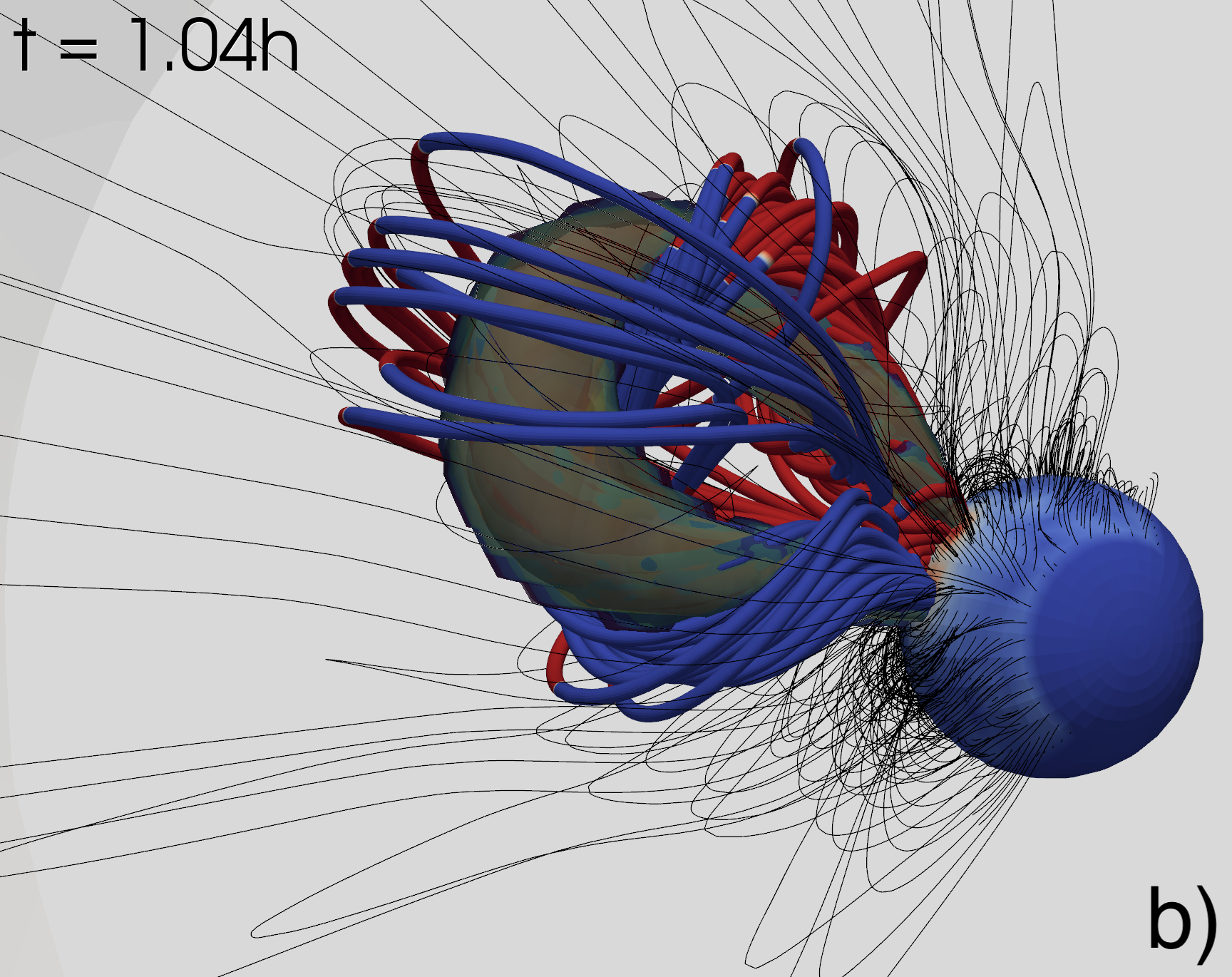}} \\
    \subfigure{\includegraphics[width=0.40\textwidth]{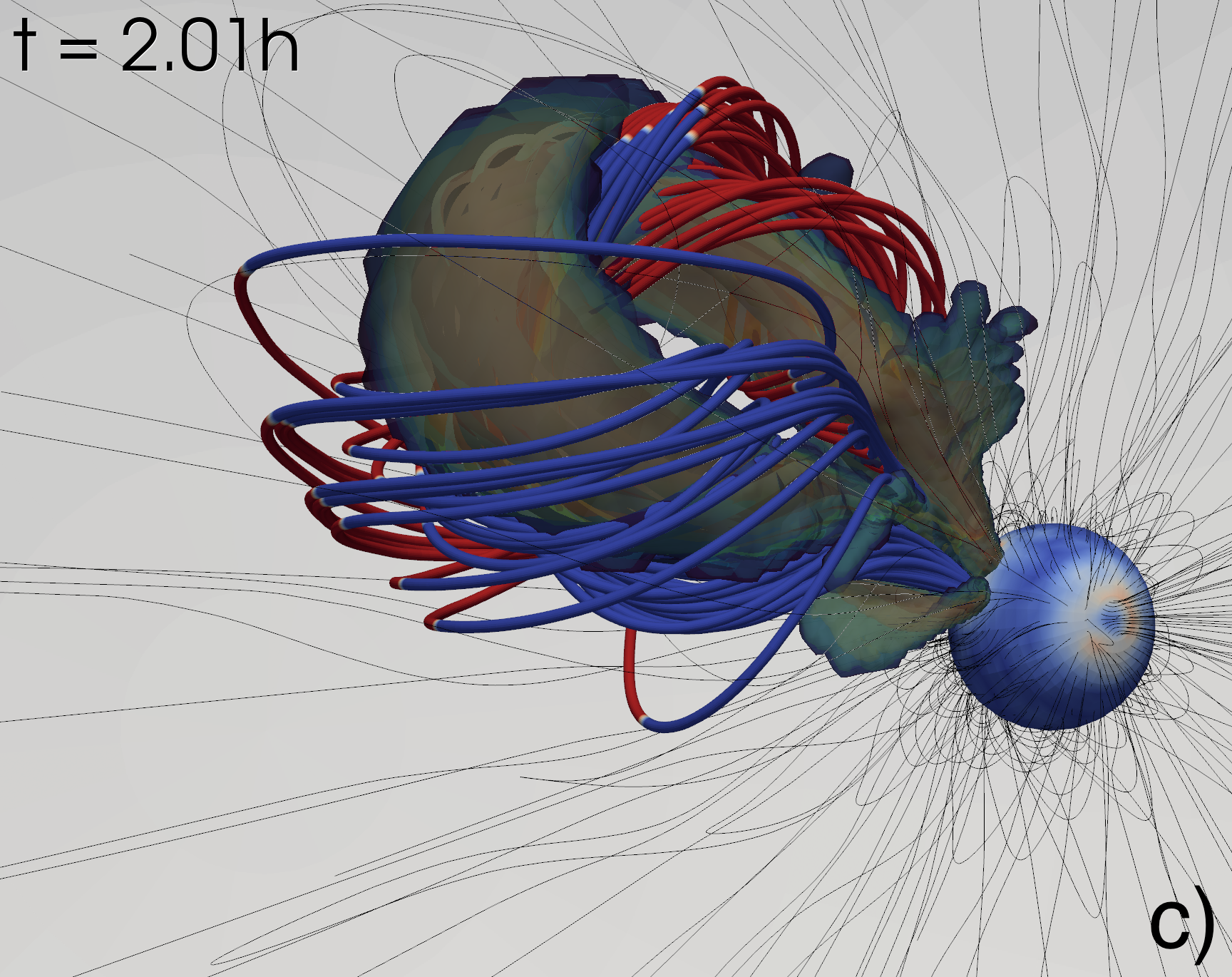}} \quad
    \subfigure{\includegraphics[width=0.40\textwidth]{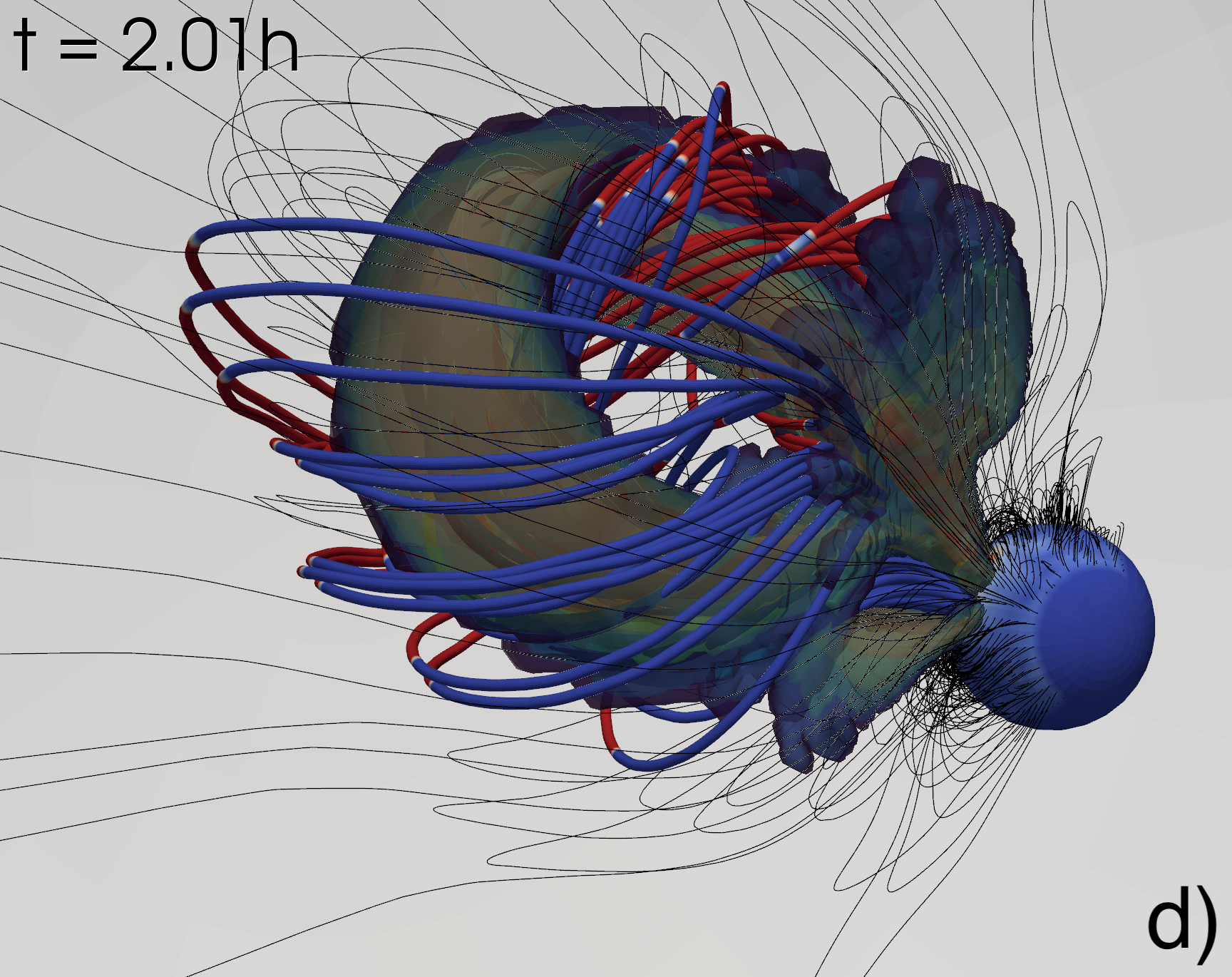}}
    \begin{minipage}[b]{0.8\textwidth}
        \includegraphics[width=\textwidth]{legend_int.png} 
    \end{minipage}
    \caption{Comparing results of the unstructured COCONUT grid and the interpolated structured grid at $t = 1.04$\,h (top row) and 2.01\,h (bottom row). Panels a) and c) show 3D plots of the particle intensity contours using the unstructured COCONUT grid, while panels b) and d) show the results using the interpolated structured grid.}
    \label{fig:compare_unstruct_vs_struct_grids}
\end{figure*}

As described in Sec.~\ref{app:coconut}, the unstructured grid in COCONUT required the implementation of alternative algorithms in PARADISE, such as those for tracing the particles or calculating the gradients of the magnetic field and velocity components on an unstructured grid. Because these new algorithms take significantly longer than the original methods in PARADISE, we developed a code to interpolate the unstructured grid onto an EUHFORIA-like structured grid, enabling us to utilize the faster existing methods. Figure~\ref{fig:compare_unstruct_vs_struct_grids} compares particle intensities at two different times (1.04\,h and 2.01\,h), where panels a) and c) show results based on the unstructured grid, and panels b) and d) are based on the interpolated grid. Since the TDFR erupts in the direction of $\phi = 180^\circ$, we limited the domain in the interpolated snapshots in colatitude to $\theta \in \{40^\circ, 140^\circ\}$ and $\phi \in \{90^\circ, 270^\circ\}$, which allowed us to increase the resolution of the interpolated snapshots. On the structured grid, the cells increased in length in the radial direction from about $0.00085\,R_\odot$ to about $1\,R\odot$, approximately matching the radial spacing of the original COCONUT grid towards the outer boundary, while having a more coarse grid at the inner boundary.

In comparison, the unstructured grid has at the inner boundary the smallest cell lengths with about $0.00025\,R_\odot$, which increase to $1\,R_\odot$ at the outer boundary. The cell spacing in colatitudinal and longitudinal directions was set constant, with an angular resolution of about $0.8^\circ$. The solar wind plasma variables were interpolated from the COCONUT grid to the structured grid using all locally connected cells (i.e., a group of cells that share at least one node with the local cell) in COCONUT, weighted linearly with distance. The plots at 1.04\,h (upper row) show the particle intensities about 30\,min after injection. While the intensities at this time are highly similar, at the later time of 2.01\,h (lower row), the particles in the simulation with the interpolated grid experience slightly more diffusion, possibly due to a smoothing of the solar wind values by the interpolation onto the structured grid. 

While we continue to improve PARADISE working with the unstructured grid, we will also enhance the interpolated model by implementing more elaborate interpolation schemes to reduce the additional diffusive effects. The wall time component plays a crucial role, as the aim is to eventually use COCONUT+PARADISE as a forecasting tool. A further advantage of using the original PARADISE architecture also with COCONUT is the ongoing coupling of COCONUT to EUHFORIA, allowing particle transport simulations from the Sun's surface up to 1\,au and beyond in PARADISE in a consistent manner (see also Sec.~\ref{sec:summary}).

\section{A Note on Numerical Diffusion}\label{app:num_diffusion}

\begin{figure*}[t]
    \centering
    \subfigure{\includegraphics[width=0.40\textwidth]{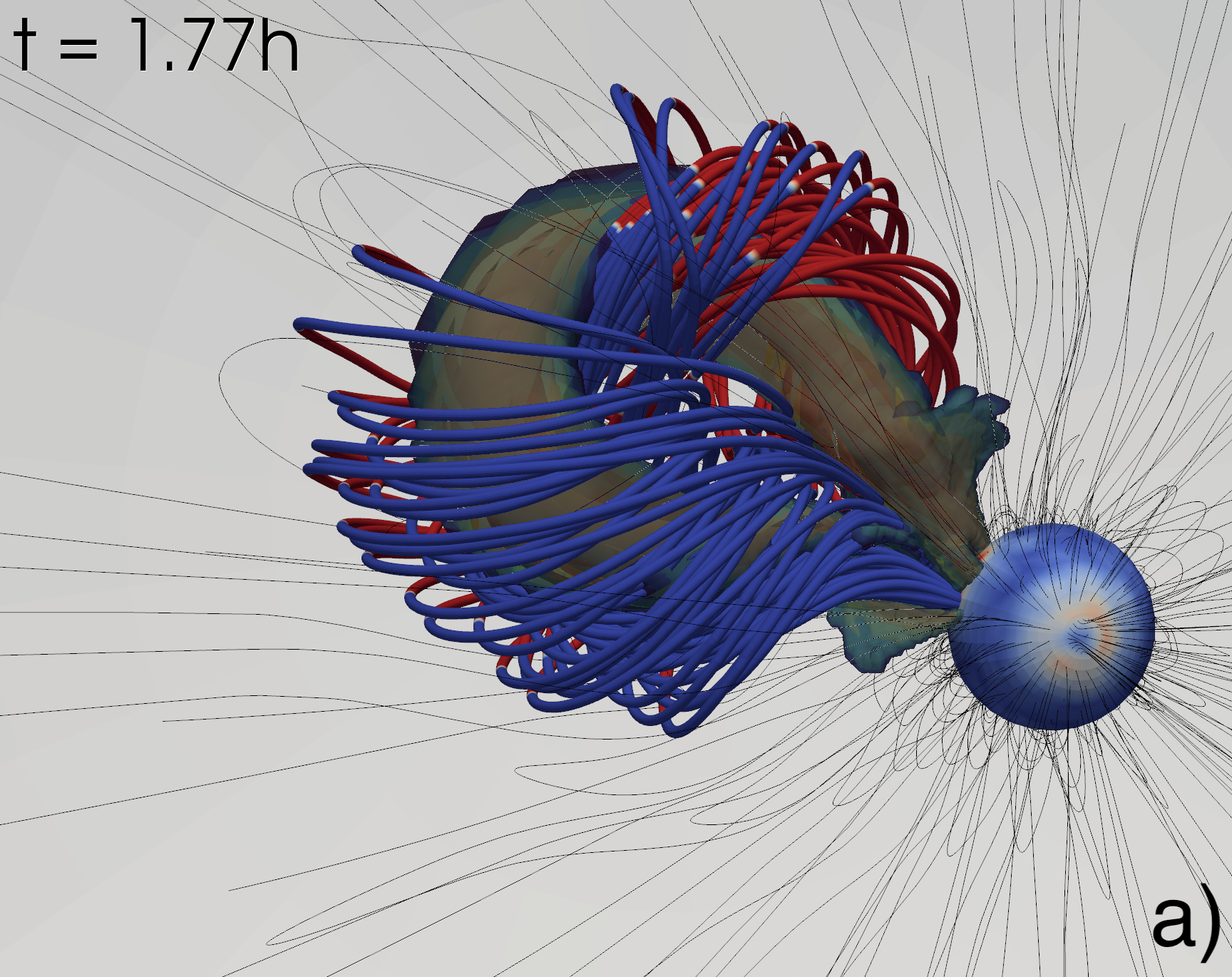}} \quad
    \subfigure{\includegraphics[width=0.40\textwidth]{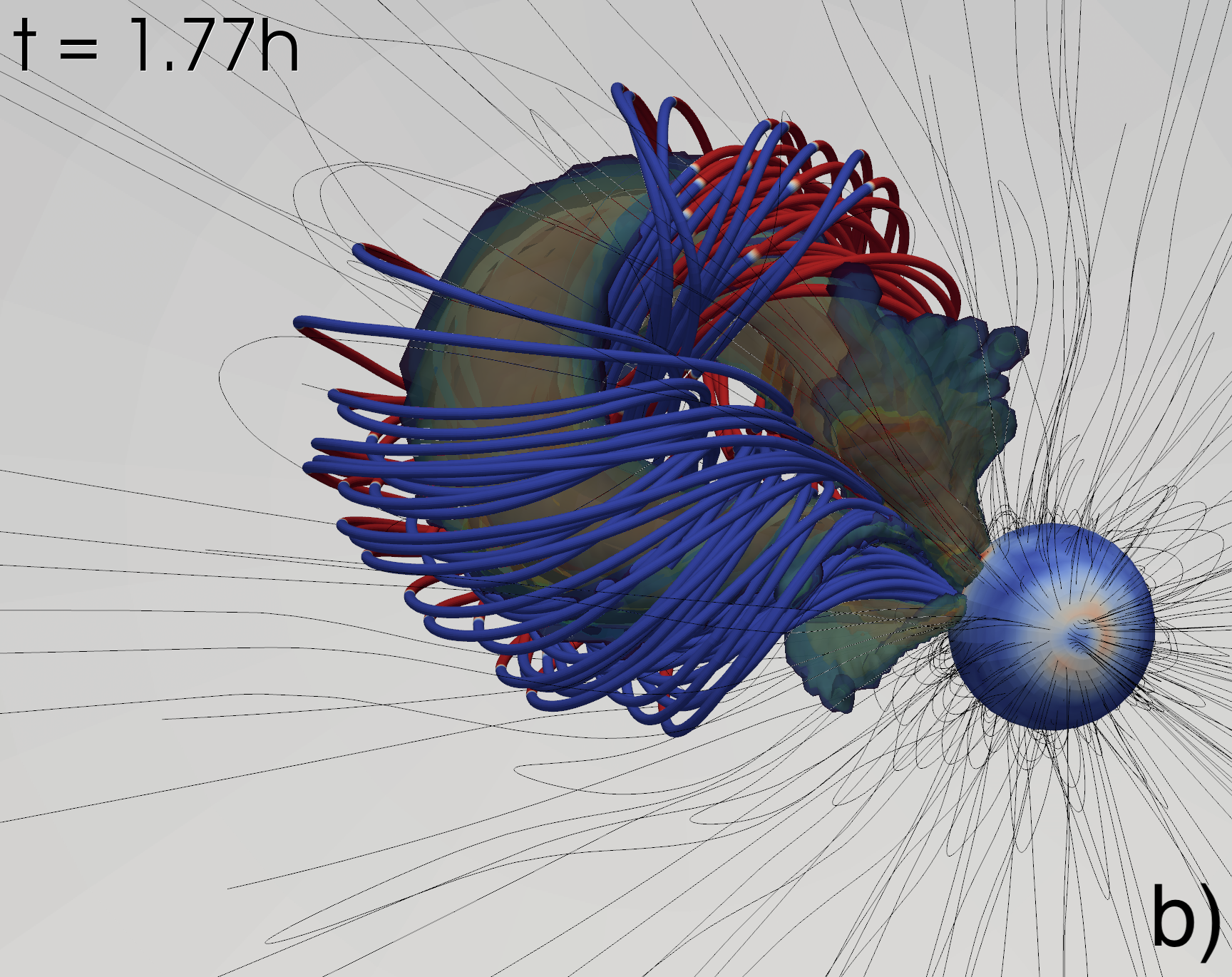}} \\
    \subfigure{\includegraphics[width=0.40\textwidth]{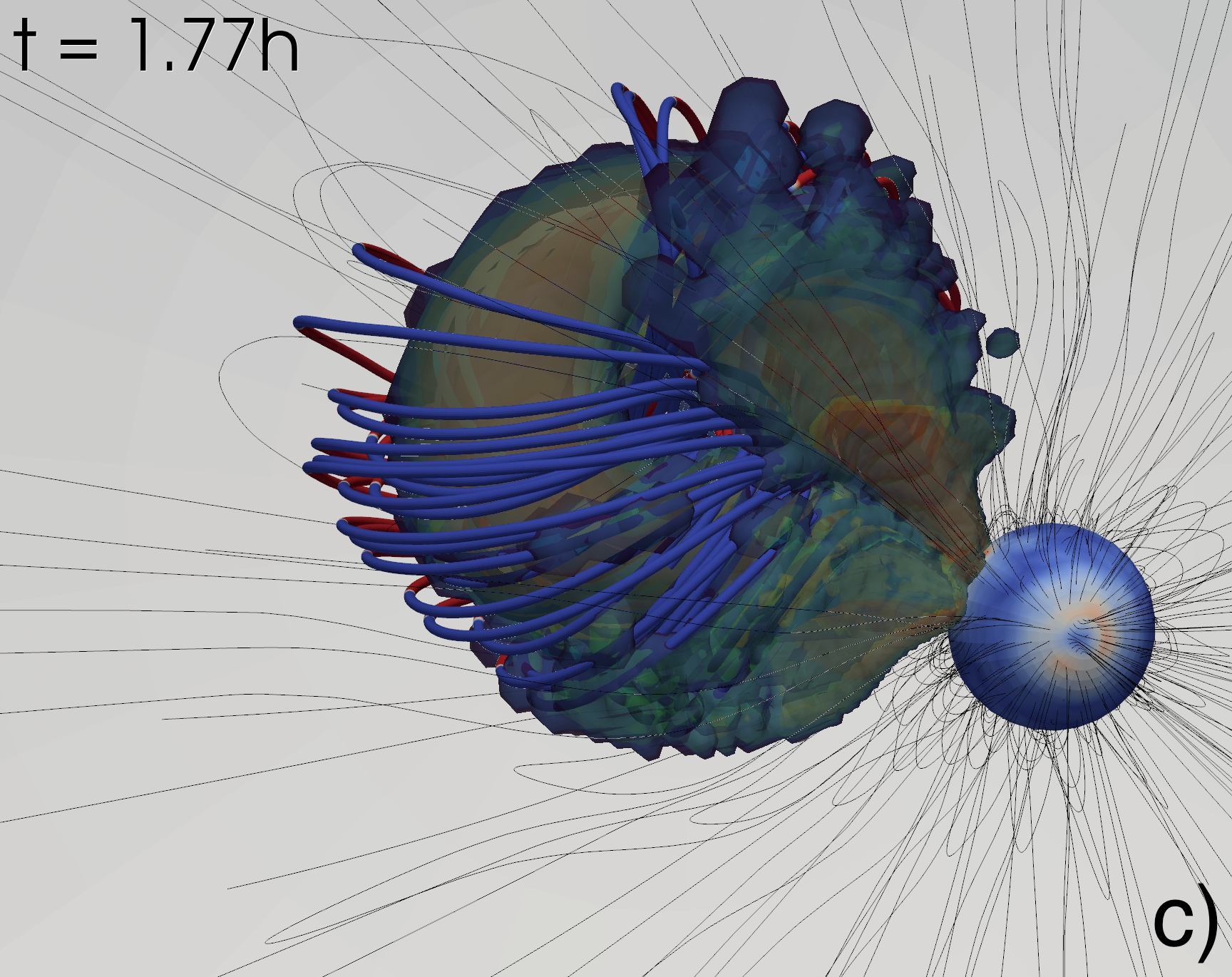}} \quad
    \subfigure{\includegraphics[width=0.40\textwidth]{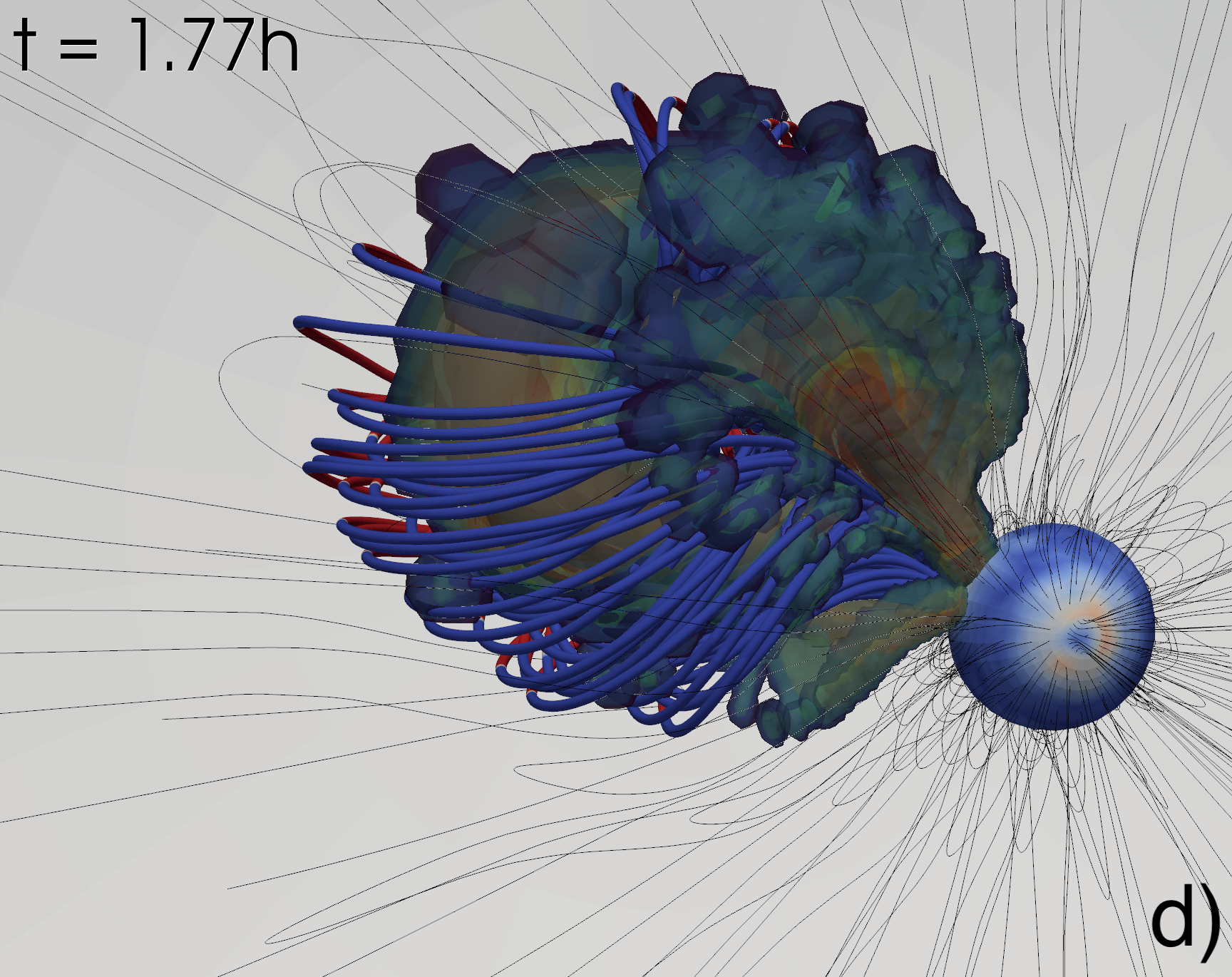}}
    \begin{minipage}[b]{0.8\textwidth}
        \includegraphics[width=\textwidth]{legend_int.png} 
    \end{minipage}
    \caption{3D particle intensity plots illustrating possible sources of numerical diffusion. Panels a) through c) are based on a COCONUT snapshot cadence of about 5\,min, 15\,min, and 20\,min, respectively, with a minimal time step of $\Delta t_\mathrm{min} = 10^{-4}$\,s in PARADISE, while in panel d) the same COCONUT snapshot cadence is used as in panel a) but with $\Delta t_\mathrm{min} = 5$\,s. The same injection samples as in Sec.~\ref{sec:results} have been used for this comparison.}
    \label{fig:num_diffusion}
\end{figure*}

To examine the effect of numerical diffusion in the particle transport simulations, we compared simulation outcomes using different timesteps within PARADISE and different output cadences in COCONUT, illustrated in Fig.~\ref{fig:num_diffusion}. The four panels show 3D plots of the particle intensities, as done in Sec.~\ref{sec:results}. Panel a) shows the results for the default setup for all simulations in the main part, employing a minimal time step of $\Delta t_\mathrm{min} = 10^{-4}$\,s and a COCONUT output cadence of about 5\,min. For panels b) and c), $\Delta t_\mathrm{min}$ is the same as in panel a), but the COCONUT output cadences are 15\,min and 20\,min, respectively. While panel b) shows only a slight increase in numerical diffusion, in panel c), the particles exhibit significant additional diffusion within the CME, causing particles to escape the CME.

In panel d), the MHD output cadence is the same as in panel a), but $\Delta t_{min}$ is increased to 5\,s. The additional numerical diffusion in panel d) is comparable to the case in panel c). Decreasing $\Delta t_{min} = 10^{-4}$\,s by a factor of 100 led to significantly longer wall times without a sufficiently notable reduction in numerical diffusion. We conclude that the minimal timestep in panel a) already causes a sufficiently small amount of numerical diffusion for the given particle energy and MHD output cadence of 289\,s, and thus $\Delta t_{min} = 10^{-4}$\,s has been selected as the default setup for all simulations in the main part of the paper. Lowering COCONUT's output cadence to a value such as 1\,min would minimize numerical diffusion further. Still, it would increase the data volume significantly, which needs to be considered regarding the potential use of COCONUT+PARADISE as a forecasting tool.

\newpage

\bibliography{cfd_paper}{}
\bibliographystyle{aasjournal}

\end{document}